\input amstex
\input epsf
\documentstyle{amsppt}
\magnification 1000
\catcode`\@=11
\def\logo\@{}
\catcode`\@=\active  
\NoBlackBoxes
\baselineskip=22pt
\topmatter
\title Semiclassical  Focusing  NLS 
with Steplike Initial Data\endtitle
\author Spyridon Kamvissis \endauthor
\endtopmatter
\document

\define\[{\left[}%
\define\]{\right]}%
\define\({\left(}%
\define\){\right)}%
\baselineskip=20pt

Department of Applied Mathematics, University of Crete,  
Greece

\bigskip

July 17, 2009

\bigskip

ABSTRACT

\bigskip

We study the semiclassical behavior of the focusing
nonlinear Schr\"odinger equation in 1+1
dimensions under discontinuous
"barrier" initial data  and we  
describe  the violent oscillations arising in terms of theta functions.
The  construction of proofs  relies on 

(i) the analysis  of the associated
Riemann-Hilbert factorization problem  

(ii) the analysis of the resulting maximin variational 
problem for a Green's potential with external field and in particular
the proof of existence of a regular solution,
which enables the construction of the so-called g-function
transformation and hence the asymptotic deformation of
the Riemann-Hilbert factorization problem to one that is explicitly solvable
in terms of theta functions.  

In particular we show  that the
finite genus ansatz is generically  valid for all times. 

\bigskip

ACKNOWLEDGEMENT. The research leading to this article was supported by the
Max Planck Society of Germany and the European Science Foundation (MISGAM program).

\newpage

0. INTRODUCTION

\bigskip

The semiclassical limit of the 
1+1-dimensional, integrable nonlinear
Schr\"odinger equation with cubic focusing nonlinearity has been the subject
of recent investigations. Several numerical studies have appeared
since 1998 ([MK], [BK], [CT], [C]) and   rigorous analyses 
of the initial value problem under real analytic data have already  
appeared ( [KMM] in 2000 and [TVZ] in 2004) in cases where the initial data is analytic.
The present paper
makes use of the method and results in [KMM] for the study of a very particular
problem with discontinuous (barrier) data. It has been shown by A.Cohen and
T.Kappeler [CK] that weak solutions exist for all time under barrier data and, even more,
that the inverse scattering technique is still applicable to the integration
of the problem. In fact, it is known ([DZ2]) that
there exists a unique solution in 
$C(\Bbb R_+, L^2(dx) \cup  (L^{\infty}(dx) \otimes   L^4_{loc}(dt))$.
Here, we use the inverse scattering method of [CK] to pose an
associated Riemann-Hilbert factorization problem, which we then asymptotically
(as $h \to 0$) reduce to the Riemann-Hilbert factorization problem
that  can be explicilty solved in terms of theta functions.

The real aim of this paper is to indicate that a discontinuity in the initial data
does not necessarily alter the behavior of the semiclassical focusing NLS problem.
Some changes of course $have$ to happen. For example, if the Euler system
that appears as a formal limit of the focusing  semiclassical NLS
does not even admit a solution for small times, it is obvious that the genus zero
ansatz cannot hold uniformly for small times. 

A natural  generalization of the barrier data problem is the problem of
general step data. For such data, the eigenvalue density
$\rho^0$ has several nonanalyticity points on the
"spike" where 
the eigenvalues accumulate to the complex plane. The analysis
of the present paper  can be immediately extended to that case.
The appropriate contour
still exists and the finite  genus ansatz still applies. The proof of that fact is
not very different from the proof of the simplest case that we present here.

\newpage

1. FOCUSING NLS WITH BARRIER DATA

\bigskip

We consider the  nonlinear
Schr\"odinger equation (1+1-dimensional, integrable, focusing case), 
on the half-line
$$
\aligned
i h u^h_t (x,t) +  {h^2 \over 2 }  u^h_{xx}(x,t)  + |u^h(x,t)|^2 u^h(x,t) = 0, \\
u^h(x,0) = u_0(x),
\endaligned
\tag 1
$$
under barrier-like initial data:
$$
\aligned
u_0(x) = A, ~~~-1/2 < x < 1/2,\\
=0, ~~~ otherwise.
\endaligned
\tag 2
$$

Here $A$ is a fixed positive constant and 
$h$ is a small positive constant.
Eventually we will take $h \to 0$. We will assume that $h$ is 
staying away from 
the discrete set $ \{ {{2A} \over {(2k+1) \pi}} \}, k=0,1,2,...$.
For simplicity, we will actually require that 

$$
\aligned
h~ takes~values~in~
the~discrete~set ~ \{ {{A} \over {k \pi}} \},~ k=0,1,2,...
\endaligned
\tag2a
$$ 

Setting 
$$
\aligned
\rho^h= |u^h|^2,\\
\mu^h = -i {h \over 2} (\bar u^h  u^h_x -  u^h  \bar u^h_x ),
\endaligned
$$
(1) is transformed to
$$
\aligned
\partial_t \rho^h +  \partial_x \mu^h =0, \\
\partial_t \mu^h + \partial_x ({ {(\mu^h)^2  }\over \rho^h}) - \partial_x (\rho^h)^2/2=
{{h^2}\over 4} \partial_x (\rho^h \partial_x^2 log \rho^h).
\endaligned
$$

The formal limit as $h \to 0$ is the Euler system
$$
\aligned
\partial_t \rho +  \partial_x \mu =0, \\
\partial_t \mu + \partial_x ({ {\mu^2  }\over \rho}) 
- \partial_x (\rho)^2/2 = 0.
\endaligned
\tag1a
$$
The initial data become $\rho =u^2_0(x), \mu=0$.

This initial value problem admits a weak solution for all time,
as shown by A.Cohen and T.Kappeler ([CK]). Furthermore, 
the inverse scattering theory is still applicable.

The associated linear system is
$$
\aligned
h \psi_x = \pmatrix & -i \lambda & u(x) \\
&  - u^*(x) &i \lambda \endpmatrix \psi,
\endaligned
\tag3
$$
where * denotes complex conjugation. Jost functions $\phi, \psi$
are defined as column vector solutions of (3)  satisfying the asymptotic
conditions
$$
\aligned
\psi \sim \pmatrix & 0 \\ & e^{i \lambda x/h} \endpmatrix , 
~as~ x \to +\infty,\\
\phi \sim \pmatrix  & e^{-i \lambda x/h} \\&0 \endpmatrix , ~as~ x \to -\infty.\\
\endaligned
\tag4
$$

One can define the reflection and transmission coefficient as follows.

Let $\psi^0$ be the transpose of $(\psi_2^*, -\psi_1^*)$.
Then, following [CK] define $a_+, b_+$ by
$$
\aligned
\phi = a_+ \psi^0 + b_+ \psi.
\endaligned
$$
The reflection and transmission coefficients are given by
$$
\aligned
r_+ =b_+ / a_+,\\
t_+ = 1/a_+.
\endaligned
$$
Similarly, one can define coefficients $r_-, t_-$ by normalizing the
Jost functions at the opposite infinities.

From [CK], we have
$$
\aligned
\psi (x, \lambda) =  \pmatrix & 0 \\ & e^{i \lambda x/h} \endpmatrix,
 ~x< -1/2,\\
=\pmatrix & A e^{i\lambda/h} ((x-1)/h) 
{{sin[(A^2+\lambda^2)^{1/2}((x-1)/h)]} \over {(A^2+\lambda^2)^{1/2} ((x-1)/h)}}
\\ & e^{i\lambda/h} cos[(A^2+\lambda^2)^{1/2}((x-1)/h)]
+\lambda  e^{i\lambda/h} {{sin[(A^2+\lambda^2)^{1/2}((x-1)/h)]} \over 
{(A^2+\lambda^2)^{1/2} ((x-1)/h) }}  \endpmatrix  ,~~~ -1/2< x < 1/2,\\
=  \pmatrix  & \beta (\lambda, h) e^{i \lambda x/h} \\  
& \alpha (\lambda, h) e^{i \lambda x/h} 
\endpmatrix,~x>1/2,
\endaligned
\tag5
$$
where
$$
\aligned
\alpha = &e^{i\lambda/h} ~(~cos[(A^2+\lambda^2)^{1/2}/h]
-i \lambda  {{sin[(A^2+\lambda^2)^{1/2}(1/h)]} \over 
{(A^2+\lambda^2)^{1/2} }}~) , \\
\beta =&-Ae^{i\lambda/h}{{sin[(A^2+\lambda^2)^{1/2}(1/h)}] \over 
{(A^2+\lambda^2)^{1/2} }}.
\endaligned
\tag6
$$
The coefficients $a_+, b_+, r$ are given by
$$
\aligned
a_+ = {{\alpha} \over {|\alpha|^2 + |\beta|^2}},\\
b_+ = {{\beta} \over {|\alpha|^2 + |\beta|^2}},\\
r(\lambda) = {{\beta} \over{\alpha}}= 
 -A {1\over {cot[(A^2+\lambda^2)^{1/2}(1/h)] (A^2+\lambda^2)^{1/2}-i\lambda  }}.
\endaligned
\tag7
$$

The eigenvalues are the zeros of $a_+$. 
They lie on the imaginary interval $[-iA,iA]$ and are given  by
$$
\aligned
\lambda = i \eta,\\tan [ {{(A^2-\eta^2)^{1/2}} \over h}]
= {{(A^2-\eta^2)^{1/2}} \over \eta}.
\endaligned
\tag8
$$
In other words,
$$
\aligned
{{(A^2-\eta^2)^{1/2}} \over h} = arctan [{{(A^2-\eta^2)^{1/2}} \over \eta}]
+ k \pi,~~~k \in \Bbb Z.
\endaligned
$$
As $h \to 0$, we get an asymptotic expression for the
eigenvalues $\lambda_k = i\eta_k$.
$$
\aligned
{(A^2-\eta_k^2)^{1/2}} \sim h k \pi,~~~k \in \Bbb Z.
\endaligned
\tag8'
$$
The limiting density of eigenvalues is then given by
$$
\aligned
{{\eta } \over{\pi (A^2-\eta^2)^{1/2} }}.
\endaligned
\tag9
$$
or, in terms of $\lambda = i \eta$,
$$
\aligned
\rho^0 (\lambda) = {{\lambda } \over{\pi   (A^2+\lambda^2)^{1/2} }}.
\endaligned
\tag9a
$$ 
Here the branch is chosen such that 
$\rho^0 (\lambda) \sim {i \over {\pi A}}$ as $\lambda \to i \infty$.

We note here that as a consequence of the
simplifying condition (2a) we have
$\int_0^{iA} \rho^0 (\lambda) d\lambda= -i A /\pi = ikh$, for some
integer $k$. Hence
$$
\aligned
exp [{{ \pi} \over h} \int_0^{iA} \rho^0 (\lambda) d\lambda] =1.
\endaligned
\tag9b
$$
This will simplify the analysis of the "local" Riemann-Hilbert problem on a cross
centered at the origin; see section 5, paragraph 6.
Another consequence of (2a) is that $r(0)=0$.

It is easily seen that the associated norming constants can only take the values
$-1, 1$ (by symmetry in $x$) and that in fact they have to oscillate
between these two values (by a Sturm-Liouville oscillation argument; see [KMM]).

REMARK.   The density $\rho^0$ blows up at $iA$. This is due to the local 
"flatness" and not the discontinuity
of the initial data. There are some minor implications of this fact. The external field
of the associated variational problem (see later) is still continuous ar $iA$ and analytic nearby.
On the other hand we do not want to allow that the band endpoints $\lambda_j$ (defined later) are equal
to $iA$. Generically (in $x,t$) this will not happen. 
If, say, $\lambda_0=iA$, then the anlysis of [KMM] breaks down. In particular it is
not clear how to solve the scalar Riemann-Hilbert problem for the g-function.

We end this section by defining the three Pauli matrices; we will be using them later.
$$
\aligned
\sigma_1 =
\pmatrix &0 &1\\&1 &0 \endpmatrix,~~~~
\sigma_2 = \pmatrix &0 &-i \\&i &0 \endpmatrix,~~~~~
\sigma_3 = \pmatrix &1 &0 \\ &0 &-1 \endpmatrix.
\endaligned
$$

\bigskip

2. THE RIEMANN-HILBERT PROBLEM

\bigskip

We can now state the following Riemann-Hilbert problem, following
[KMM].
Let $C$ be a piecewise smooth loop encircling all eigenvalues in
the upper half-plane and lying entirely in
the upper half-plane except for the point $0$.
Let   $C^*$ be its conjugate, encircling the eigenvalues
in the lower half-plane. 
Also give the following orientation:

(i) the real axis is oriented  from left to right,

(ii) the loop $C$ is oriented counterclockwise,

(iii) the loop $C^*$ is also oriented  counterclockwise.

We use the following convention: the +-side of an oriented 
contour is always to its left, according to the given orientation.

We also choose $C$ so that it  approaches $0$ (from left and right) at a non-zero, non-straight
angle with the real axis.

THEOREM 1 (discrete version). Let $d\mu = \Sigma_k h (\delta_{\lambda_k^*} - \delta_{\lambda_k})$
be a finite sum of point  measures supported at the eigenvalues of the system (3), 
as given by (8).  Let 
$$
\aligned
X (\lambda) = -  (A^2+\lambda^2)^{1/2}.
\endaligned
\tag10
$$

Letting $M_+ $ and $M_-$ denote the limits of 
$M$  on $\Sigma = C \cup C^*$ from  left and right 
respectively, we define the Riemann-Hilbert factorization problem 
$$
\aligned
M_+(\lambda) = M_-(\lambda) J(\lambda) ,\\
where~~~\\J(\lambda) = 
\pmatrix
&1  &r(\lambda)  e^{{-2i \lambda x - 2i\lambda^2 t} \over h} \\ 
&r^* (\lambda) e^{{2i \lambda x + 2i\lambda^2 t} \over h}&  1+|r(\lambda)|^2
\endpmatrix, 
\lambda \in \Bbb R,\\
=v(\lambda), \lambda \in C,\\
= \sigma_2 v(\lambda^*)^* \sigma_2, \lambda \in C^*,\\
lim_{\lambda \to \infty} M(\lambda) =I,
\endaligned
\tag11
$$
where
$$
\aligned
v(\lambda)=
\pmatrix
&1 &  -i~exp ({1 \over h} \int log(\lambda -\eta) d\mu(\eta))
exp (-{1 \over h} (2i\lambda x + 2i \lambda^2 t - X(\lambda)))\\
&0 &1  
\endpmatrix,
\endaligned
\tag12
$$
and $r$ is the reflection coefficient  defined in (7):
$$
\aligned
r(\lambda) = -A {1\over 
{cot[(A^2+\lambda^2)^{1/2}(1/h)] (A^2+\lambda^2)^{1/2}-i\lambda  }}.
\endaligned
\tag13
$$
Note that $r(\lambda)$ can have a real singularity at $\lambda =0$ if $A/h$
is an odd multiple of $\pi/2$. By assumption, we have excluded 
such values of $h$.

The above Riemann-Hilbert problem 
admits a solution and the solution of (1) can be recovered from the solution of (11) as follows.

$$
\aligned
u(x,t) = 2i  h ~lim_{\lambda \to \infty}  (\lambda  M^{12}(\lambda)),
\endaligned
\tag14
$$
where the index 12 here denotes the (12)-entry of a matrix.

PROOF: Standard; see Chapter 2 of [KMM]. The quantity $-i~exp ({1 \over h} (X(\lambda_k)))$
oscillates between $-1$ and $1$. The function 
$-i~ exp ({1 \over h} (X(\lambda)))$
is thus an extrapolation of the norming constants.

REMARKS. 1. In [KMM] the Riemann-Hilbert problem jump involves some
parameters denoted by $K, \sigma, J$. Here we are simply choosing
$K=-1, \sigma=1, J=1$. This is compatible with the discussion in [KMM]
as long as we focus our attention to the case $x \geq 0$. By the obvious symmetry
$x \to -x$ this is acceptable.

2. Obviously, the contour $C$ can be deformed anywhere in the upper half-plane,
as long as it passes through $0$ (we only require it to be
non-tangent to either the real axis or the imaginary axis at $0$) 
and does not touch the
linear segment $[0, iA]$. For the asymptotic analysis, the contour $C$
will be eventually fixed by the choice of
the "g-function" in the next section. Similarly for its conjugate $C^*$.
In a sense we are seeking a $steepest ~descent~ contour$, on which the 
Riemann-Hilbert problem will reduce to one that is $explicitly~solvable$.

\bigskip
 
We note the following factorization of the jump $J(\lambda)$ on the real
line.
$$
\aligned
\pmatrix
&1  &r(\lambda)  e^{{-2i \lambda x - 2i\lambda^2 t} \over h} \\ 
&r^* (\lambda) e^{{2i \lambda x + 2i\lambda^2 t} \over h}&  1+|r(\lambda)|^2
\endpmatrix = L(\lambda) U(\lambda),~~~\\where\\
~~L(\lambda)=
\pmatrix
&1  &0 \\ 
&r^* (\lambda^*) e^{{2i \lambda x + 2i\lambda^2 t} \over h}&  1
\endpmatrix,~~~~~~
U(\lambda) =
\pmatrix
&1  &r(\lambda)  e^{{-2i \lambda x - 2i\lambda^2 t} \over h} \\ 
&0&  1 \endpmatrix.
\endaligned
\tag15+
$$

A different factorization is also possible.
$$
\aligned
\pmatrix
&1  &r(\lambda)  e^{{-2i \lambda x - 2i\lambda^2 t} \over h} \\
&r^* (\lambda) e^{{2i \lambda x + 2i\lambda^2 t} \over h}&  1+|r(\lambda)|^2
\endpmatrix =  S(\lambda) D(\lambda) T(\lambda),\\ ~~~~where\\
 S(\lambda)=\pmatrix
&1  &{{r(\lambda)} \over{ 1+|r(\lambda)|^2}}  e^{{-2i \lambda x - 2i\lambda^2 t} \over h} \\
&0&  1 \endpmatrix.
\\D(\lambda)= \pmatrix
& (1+|r(\lambda)|^2)^{-1}   &0 \\
&0& 1+|r(\lambda)|^2 \endpmatrix,\\
T(\lambda)= \pmatrix 
&1   &0 \\
&{{r^* (\lambda^*)}  \over{ 1+|r(\lambda)|^2}}  e^{{2i \lambda x + 2i\lambda^2 t} \over h}& 1 
\endpmatrix.  
\endaligned
\tag15-
$$

Both factorizations will be useful later.
 
\bigskip

REMARK: From now on we will substitute 
$ d\mu = (\rho^0 (\eta) + (\rho^0)^* (\eta^*)) d\eta$ in (12),
where $\rho^0$ is the asymptotic density of eigenvalues given by (9a),
and supported on the linear segment $[0,iA]$. In
other words, we will approximate a discrete density of eigenvalues by
a continuous one. This is not a trivial assumption, but it is true.
We refer to  Chapter 3 of [KMM] for a rigorous justification.
Of course the justification of [KMM] is devised for real analytic data,
but the proof goes through unaltered since the nice behavior of 
the asymptotic density of eigenvalues at the crucial point $0$ is not altered 
even when we consider steplike data.

We then restate Theorem 1 as follows.

THEOREM 1 (continuous version). Let 
$ d\mu = (\rho^0 (\eta) + (\rho^0)^* (\eta^*)) d\eta$,
where $\rho^0$ is the asymptotic density of eigenvalues given by (9a),
and supported on the linear segment $[0,iA]$. Set
$$
\aligned
X (\lambda) = -  (A^2+\lambda^2)^{1/2}.
\endaligned
\tag10
$$

Letting $M_+ $ and $M_-$ denote the limits of
$M$  on $\Sigma = C \cup C^*$ from  left and right
respectively, we define the Riemann-Hilbert factorization problem
$$
\aligned
M_+(\lambda) = M_-(\lambda) J(\lambda) ,\\
where~~~\\J(\lambda) =
\pmatrix
&1  &r(\lambda)  e^{{-2i \lambda x - 2i\lambda^2 t} \over h} \\
&r^* (\lambda) e^{{2i \lambda x + 2i\lambda^2 t} \over h}&  1+|r(\lambda)|^2
\endpmatrix,
\lambda \in \Bbb R,\\
=v(\lambda), \lambda \in C,\\
= \sigma_2 v(\lambda^*)^* \sigma_2, \lambda \in C^*,\\
lim_{\lambda \to \infty} M(\lambda) =I,
\endaligned
\tag11'
$$
where
$$
\aligned
v(\lambda)=
\pmatrix
&1 &  -i~exp ({1 \over h} \int log(\lambda -\eta) d\mu(\eta))
exp (-{1 \over h} (2i\lambda x + 2i \lambda^2 t - X(\lambda)))\\
&0 &1
\endpmatrix,
\endaligned
\tag12'
$$
$$
\aligned
r(\lambda) = -A {1\over
{cot[(A^2+\lambda^2)^{1/2}(1/h)] (A^2+\lambda^2)^{1/2}-i\lambda  }}.
\endaligned
\tag13
$$

The above Riemann-Hilbert problem
admits a solution and the solution of (1) can be 
asymptotically recovered from the solution of (11) as follows.
As $h \to 0$, 
$$
\aligned
u(x,t) \sim  2i  h ~lim_{\lambda \to \infty}  (\lambda  M^{12}(\lambda)),
\endaligned
\tag14'
$$
where the index 12 here denotes the (12)-entry of a matrix.

PROOF: See Chapter 3 of [KMM] in the case where the maximizing contour of the
variational problem (see Appendix 2) does not touch the spike $[0,iA]$. 
For the general case, follow the discussion in Appendix 2.

\bigskip

3. ASYMPTOTIC ANALYSIS OF THE RIEMANN-HILBERT PROBLEM.

\bigskip
The ideas underlying the discussion of this section are the following.

(i) Our given Riemann-Hilbert problem needs to be asymptoticaly deformed
to an explicltly solvable one. To do this, some analyticity properties
of the jump matrix are crucial. Proofs of the passage from one  
Riemann-Hilbert problem to another rely heavily 
on the equivalence of  Riemann-Hilbert problems
and special singular integral equations (see [DZ1]
for the first rigorous description of the 
deformation method).

(ii) One of the crucial deformations involves a certain conjugation
and the introduction of an appropriate "g-function" (first introduced
in [DVZ1]; see also [DVZ2]). See (17) below.

(iii) For the focusing nonlinear Schr\"odinger problem and
because the associated Lax operator is $not ~ self-adjoint$ the 
appropriate introduction of the g-function requires finding
a "steepest descent contour"; see (18) below. The term "nonlinear steepest descent 
method" often applied to the Riemann-Hilbert deformation theory initiated in
[DZ1] thus acquires full meaning. For this we follow the ideas of
[KMM], [KR]; see  also Appendix A.2.

\bigskip

We begin with the following observation. Consider the reflection coefficient
given by (13). For $\lambda$ in the upper half-plane, at least away from the real
line and the eigenvalues given by (8), we have
$$
\aligned
r(\lambda) \sim {{-iA} \over { (A^2+\lambda^2)^{1/2} + \lambda}}.
\endaligned
\tag16
$$
while for $\lambda$ in the lower half-plane, at least away from the real
line and the eigenvalues given by (8),  we have
$$
\aligned 
r(\lambda) \sim {{iA} \over { (A^2+\lambda^2)^{1/2} - \lambda}}. 
\endaligned 
\tag16*
$$ 

This  means that
$r(\lambda)  e^{{-2i \lambda x - 2i\lambda^2 t} \over h}$
is exponentially decaying (growing) in the upper half-plane, 
at least away from the real and imaginary axes, as long
as $Re (\lambda) < -x/2t$  ($Re (\lambda) > -x/2t$),
while $r^*(\lambda^*)  e^{{2i \lambda x + 2i\lambda^2 t} \over h}$ 
is exponentially decaying  (growing) in the lower half-plane,
at least away from the real and imaginary axes,
as long as  $Re (\lambda) < -x/2t$ ($Re (\lambda) > -x/2t$).

The above suggests that using the factorizations of the Riemann-Hilbert problem  
defined by (15-), (15+) and applying the obviously suggested
deformations, the jump across the real line should be
reduced to a diagonal matrix independent of $x,t$.
 
To do this of course, we must ensure that the solution of the
Riemann-Hilbert problem
is uniformly bounded (in $x,t$) as
$h \to 0$. This is actually not true as will be seen later.
What is true however, is that a judicious
conjugation of the jump matrix will "deform" it to a new Riemann-Hilbert
problem whose solution $is$  uniformly bounded (in $x,t$) as 
$h \to 0$. At $that$ point we can neglect $all$ terms involving $r$.
Furthermore, the very same conjugation
will ensure that the new Riemann-Hilbert
is explicitly solvable (asymptotically).
The "deformed" Riemann-Hilbert
problem will be defined with respect to  a $steepest~descent~contour~C$.
 
More precisely,  we introduce the change of variables
$$
\aligned
Q(\lambda) = M(\lambda) e^{{g \sigma_3} \over {h}},
\endaligned
\tag17
$$
where $\sigma_3 = diag(1, -1)$ (a Pauli matrix) and the 
complex-valued function
$g$ is  constrained 
by the following conditions:
$$
\aligned
g(\lambda) ~is ~independent ~of ~h.\\
g(\lambda) ~is ~~analytic ~~~for
\lambda\in \Bbb C \setminus (C\cup C^*).\\
g(\lambda)\rightarrow 0 ~~as~~
\lambda\rightarrow\infty.\\
g(\lambda) ~~~assumes~~
continuous ~~boundary~~
values ~~from ~~both ~~~sides ~~~~of ~~~C\cup C^*,\\
denoted~~by~~~g_+(g_-)~~on~~the~~left~~(right)~~~of~~~C \cup C^*.\\
g(\lambda^*)+g(\lambda)^* = 0 ~~~~~for~~~
all ~\lambda\in \Bbb C\setminus (C\cup C^*). 
\endaligned
\tag17a
$$
These conditions of course do not define $g$ uniquely. They will be augmented by two conditons below, 
that will also implicitly define an admissible $steepest~descent~contour~C$.

The assumptions above permit us to write $g$ in terms of a measure $\rho$ defined on the
contour $C$. Indeed
$$
\aligned
g(\lambda) = {1 \over 2} \int_{C \cup C^*} log(\lambda -\eta) \rho(\eta) d\eta,
\endaligned
\tag18
$$
for an appropriate definiton of the logarithm branch (and
for $x >0$; if $x <0$ there is a sign change in [KMM] but of course
we can restrict ourselves to the case $x>0$ here because of the symmetry of NLS).

For $\lambda\in C$, define
the functions
$$
\aligned
\theta(\lambda):=i(g_+(\lambda)-
g_-(\lambda)),\\
\phi(\lambda)
:=
\int_{0}^{iA} log(\lambda-\eta)\rho^0(\eta)\,d\eta  +
\int_{-iA}^0 log(\lambda-\eta)\rho^0(\eta^*)^*\,d\eta \\
+ 
2i\lambda x + 2i\lambda^2 t + i\pi \int_\lambda^{iA}\rho^0(\eta)\,d\eta - 
g_+(\lambda) - g_-(\lambda).
\endaligned
\tag18a
$$

Now we spell out the two important conditions which  determine 
the (steepest descent) contour $C$ (not uniquely) and the function $g$. 

$$
\aligned
\rho (\eta) d\eta ~~is~~a~~real~~~measure ,\\
\Re(\phi(\lambda)) \leq 0,
\endaligned
\tag18b
$$
i.e.
a "measure reality" condition  and what can be interpreted
(see Chapter 8 of [KMM]) as a "variational
inequality" condition.
In fact, one can eventually
show that the measure $\rho(\eta) d\eta$ has to be
nonpositive: strictly zero in the "gaps" and strictly negative in the 
"bands"; see section 4 later.

\bigskip

REMARKS. 1. The fact that there exist a contour $C$ and a function $g$ 
defined implicitly by (17), (17a), (18), (18a), (18b) is highly non-trivial.
In [KMM] we had not proved this fact for all times. Instead it was only proved for small 
(finite) times and it was simply assumed for larger times. A result for larger times
appeared only later in [KR] (see Appendix 2 for a discussion). Having the results of
[KR] (or more accurately a modification of them as in Appendix 2) we do not need 
to dwell too much in the details of the construction of $C, g$ any more.

2. The function $g$ and the steepest descent
contour $C$  depend only on the density
of the eigenvalues, $\rho^0$. They are independent of the reflection coefficient $r$.

\bigskip

By (17) the Riemann-Hilbert problem for $Q$ is 
$Q_+(\lambda) = Q_-(\lambda) v_Q(\lambda),$
where
$$
\aligned
v_Q(\lambda)=
\pmatrix
&1 & r(\lambda) e^{{2g -2i \lambda x - 2i\lambda^2 t} \over h} \\
& r^*(\lambda^*) e^{{-2g +2i \lambda x + 2i\lambda^2 t} \over h}
&1+|r(\lambda)|^2 
\endpmatrix,~~~~~ \lambda \in \Bbb R,\\
v_Q(\lambda)= \pmatrix
&e^{{g_+ - g_-}\over h}
&  -i~exp ({1 \over h} \int log(\lambda -\eta) d\mu(\eta))
exp (-{1 \over h} (-g_+-g_- +2i\lambda x + 2i \lambda^2 t - X(\lambda)))\\
&0 &e^{{g_- - g_+}\over h}
\endpmatrix, \\
for~~~\lambda \in C,\\
v_Q(\lambda)= \pmatrix
&e^{{g_+ - g_-}\over h}
&0\\
&  i~exp ({-1 \over h} \int log(\lambda -\eta) d\mu(\eta))
exp ( {1 \over h} (-g_+-g_- +2i\lambda x + 2i \lambda^2 t + X^*(\lambda^*)))
&e^{{g_- - g_+}\over h} \endpmatrix, 
\\for~~~~\lambda \in C^*.
\endaligned
$$
Also $lim_{\lambda \to \infty} Q(\lambda) =I.$

$Now$, we $will$ be able to treat the terms involving the reflection coefficient
$r$ by arguing as follows.
First we note that $g$ is purely imaginary for real $\lambda$; this
follows from (18)-(18b). So, write
$g = i\psi$, so that  $\psi(\lambda) \in \Bbb R$ for $ \lambda \in \Bbb R$.
Let $\zeta(x,t, \lambda) = \psi-  \lambda x - \lambda^2 t$. Clearly,
$\zeta(\lambda) \in \Bbb R$ for $ \lambda \in \Bbb R$.

Next, divide the real line into a union of 
(finitely many) intervals, say $J_k$,
such that for $ \lambda \in interior(J_k)$, either ${{d\zeta} \over {d\lambda}} >0$ or
${{d\zeta} \over {d\lambda}} <0$. Denote by $J_l^+$ the intervals in which
${{d\zeta} \over {d\lambda}} >0$ and by   $J_l^-$ the intervals in which
${{d\zeta} \over {d\lambda}} <0$.
Naturally, the (nonzero) 
endpoints of the intervals $J_k$ are given by the condition
${{d\zeta} \over {d\lambda}} =0$.
Note that $\zeta$ is real analytic in
$\Bbb R \setminus 0$, and $\zeta \sim - \lambda^2 t$, as $ \lambda \to \pm \infty$.
The finiteness of the number of intervals follows from the analyticity of
$\zeta$, at least away from $0$. But also near $0$, 
we will show that ${{d\zeta} \over {d\lambda}} <0$.

We first show that when $\lambda =0$, then
${{d\zeta} \over {d\lambda}} = -\infty$. 
Recall that 
$$
\aligned
g(\lambda) = {1 \over 2} \int_{C \cup C^*} log(\lambda-\eta) \rho (\eta) d\eta,
\endaligned
$$
where $C \cup C^*$ is oriented counterclockwise and $\rho (\eta) d\eta$
is a non-positive measure  on $C$ which is strictly negative in a non-trivial subset of $C$ and
is extended to $C^*$ by the condition
$$
\aligned
\rho (\eta^*)  = (\rho (\eta))^*.
\endaligned
$$

Differentiating, one gets
$$ 
\aligned 
{ {dg(\lambda) } \over {d\lambda}} = {1 \over 2} \int_{C \cup C^*} (\lambda-\eta)^{-1} 
\rho (\eta) d\eta, 
\endaligned 
$$ 
and at $\lambda =0$,
$$  
\aligned 
{ {dg(\lambda) } \over {d\lambda}} = {-1 \over 2} \int_{C \cup C^*}  
{{\rho (\eta)} \over \eta} d\eta. 
\endaligned  
$$  
Making use of the symmetry with respect to complex reflection and remembering
that the orientation is counterclockwise for both $C$ and $C^*$ one gets
$$
\aligned
[ { {dg(\lambda) } \over {d\lambda}} ] (\lambda=0)
= i \int_C {{\rho (\eta) Im(\eta) }\over {|\eta|^2}} d\eta. 
\endaligned
$$
The integrand is strictly negative and the integral diverges
because $\rho$ is nonzero at zero.
It follows that 
at $\lambda =0, ~~~{{dg} \over {d\lambda}} = -i \infty,$
so $~~~{{d\zeta} \over {d\lambda}} = -\infty$. 

Now, if $\lambda$ is real and close but not equal to $0$,
$$
\aligned
[ { {dg(\lambda) } \over {d\lambda}} ] \sim
i \int_C {{\rho (\eta) Im(\eta) }\over {|\lambda-\eta|^2}} d\eta ,
\endaligned
$$
so $~~~{{d\zeta} \over {d\lambda}} < 0$.
Hence the point $\lambda =0$ belongs to the interior of some interval
$J_l^-$.

Consider first the intervals $J_l^+$. By the Cauchy-Riemann relations 
we have $d(Im\zeta) / d(Im\lambda) >0,$ 
for $\lambda \in J_l^+$,
in the positive direction perpendicular to the real line.
This means that in an area of the upper half-plane, close to $J_l^+$,
the real part of $i\zeta$ is negative,
so $exp (i\zeta /h)$ is exponentially decaying. Similarly, in an area of the upper half-plane, 
close to $J_l^-$, the real part of $i\zeta$ is positive,
so $exp (-i\zeta /h)$ is exponentially decaying.
In the particular case of $\lambda =0$ it is also easy to check that
$exp (-i\zeta /h)$ is exponentially decaying as $\lambda $ moves upwards along the
positive imaginary axis.

\bigskip

Let us now introduce the following lens-like contours.
For each interval $J_l^+$ consider small
piecewise  linear deformations upwards, say
$J_l^{+,up}$, keeping the end points fixed, but otherwise lying entirely
in the upper half-plane. Similarly, consider
small
piecewise  linear deformations downwards, say
$J_l^{+,dn}$, keeping the end points fixed, but otherwise lying entirely
in the lower  half-plane.
For each interval $J_l^-$ consider small
piecewise linear deformations upwards, say
$J_l^{-,up}$, keeping the end points fixed, but otherwise lying entirely
in the upper half-plane. Similarly,  consider
small
piecewise linear deformations downwards, say
$J_l^{-,dn}$, keeping the end points fixed, but otherwise lying entirely
in the lower  half-plane.
All orientations are compatible with  $Re(\lambda)$ increasing.
We also make sure that $J_l^{+,up}, J_l^{+,dn}, J_l^{-,up}, J_l^{-,dn}$
cut the real line at angles $\neq 0, \pi/2$.
See Figure 1.

\bigskip

\epsfxsize=15truecm
\epsffile{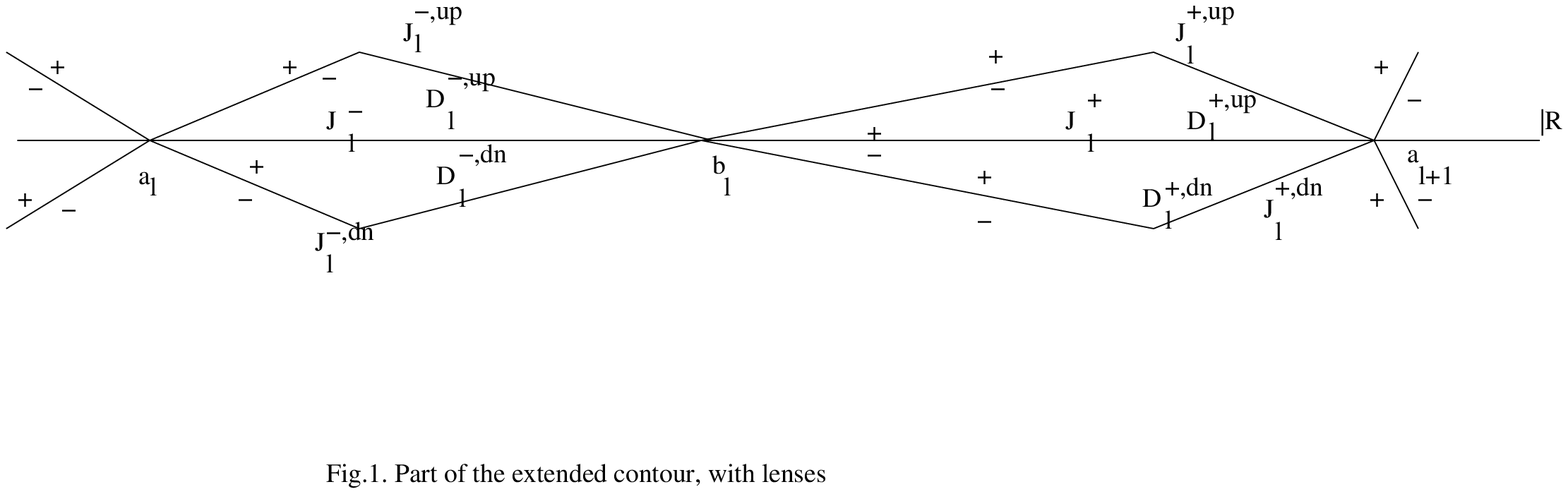}

\bigskip

Note here that one of the lens-like contours $J_l^{-,up}$ will cut the contour $C$
while one of the lens-like contours $J_l^{-,dn}$ will cut the contour $C^*$.

Let the open regions bounded by $J_l^+$ and $J_l^{+,up}$ be denoted by 
$D_l^{+,up}$ and the open regions bounded by $J_l^+$ and $J_l^{+,dn}$ be denoted by
$D_l^{+,dn}$. Similarly, let the open
regions bounded by $J_l^-$ and $J_l^{-,up}$ be denoted by  
$D_l^{-,up}$ and  the open regions bounded by $J_l^-$ and $J_l^{-,dn}$ be denoted by
$D_l^{-,dn}$.
We make sure that $Re(i\zeta) <0, $ in
$D_l^{+,up} \cup J_l^{+,up} \cup D_l^{-,dn} \cup J_l^{-,dn}$.
Similarly, we make sure that $Re(i\zeta) >0, $ in
$D_l^{-,up} \cup J_l^{-,up} \cup D_l^{+,dn} \cup J_l^{+,dn}$. 

Next, we can make use of the factorizations given in (15+), (15-).
Set
$$
\aligned
Z(\lambda) = Q(\lambda),~~~ for~~~~\lambda \in \Bbb C \setminus
\cup_l [\bar D_l^{+,up} \cup \bar D_l^{+,dn} \cup \bar D_l^{-,up} \cup 
\bar D_l^{-,dn}],\\
Z(\lambda) = Q(\lambda) U^{-1} (\lambda),~~~~for~~~~\lambda \in 
\cup_l D_l^{+,up},\\
Z(\lambda) =  Q(\lambda) L (\lambda), ~~~~for~~~~\lambda \in 
\cup_l D_l^{+,dn},\\
Z(\lambda) = Q(\lambda) T^{-1} (\lambda),~~~ for~~~~\lambda \in 
\cup_l D_l^{-,up},\\
Z(\lambda) = Q(\lambda) S (\lambda),~~~ for~~~~\lambda \in 
\cup_l D_l^{-,dn}.\\
\endaligned
\tag19
$$
The Riemann-Hilbert problem for $Z$ is described by
$$ 
\aligned 
Z_+(\lambda) = Z_-(\lambda) v_Z(\lambda),\\
v_Z(\lambda) =
\pmatrix
&1  &-r(\lambda)  e^{{2g -2i \lambda x - 2i\lambda^2 t} \over h} \\
&0&  1 \endpmatrix,~~~~for~~~~\lambda \in 
\cup_l J_l^{+,up},~~~~~~~~~\\
v_Z(\lambda) = \pmatrix
&1  &0 \\
&r^* (\lambda^*) e^{{-2g + 2i \lambda x + 2i\lambda^2 t} \over h}&  1
\endpmatrix ,~~~~for~~~~\lambda \in 
\cup_l J_l^{+,dn},\\
v_Z(\lambda) = 
\pmatrix
&1   &0 \\
& {{r^* (\lambda^*)}  \over{ 1+|r(\lambda)|^2}}  
e^{{-2g + 2i \lambda x + 2i\lambda^2 t} \over h}& 1
\endpmatrix,~~~ for~~~~\lambda \in 
\cup_l J_l^{-,up},\\
v_Z(\lambda) =  
\pmatrix
&1  &{{r(\lambda)} \over{ 1+|r(\lambda)|^2}}  e^{{2g -2i \lambda x - 2i\lambda^2 t} \over h} \\
&0&  1 \endpmatrix,~~~ for~~~~\lambda \in  \cup_l J_l^{-,dn}, \\
v_Z(\lambda)= \pmatrix
&e^{{g_+ - g_-}\over h}
&  -i~exp ({1 \over h} \int log(\lambda -\eta) d\mu(\eta))
exp (-{1 \over h} (-g_+-g_- +2i\lambda x + 2i \lambda^2 t - X(\lambda)))\\
&0 &e^{{g_- - g_+}\over h}
\endpmatrix, \\
for~~~\lambda \in C,\\
v_Z(\lambda)= \pmatrix
&e^{{g_+ - g_-}\over h}
&0\\
&  i~exp ({-1 \over h} \int log(\lambda -\eta) d\mu(\eta))
exp ( {1 \over h} (-g_+-g_-+2i\lambda x + 2i \lambda^2 t + X^*(\lambda^*)))
&e^{{g_- - g_+}\over h} \endpmatrix,
\\for~~~~\lambda \in C^*,
\\
v_Z(\lambda)= \pmatrix 
&( 1+ |r(\lambda)|^2)^{-1}   &0 \\
&0&  1+ |r(\lambda)|^2
\endpmatrix,~~~~~for~~~~\lambda \in
\cup_l J_l^-,
\endaligned
\tag20
$$
with the normalization
$lim_{\lambda \to \infty} Z(\lambda) =I.$ For a part of the extended contour,
away from 0, and between the points $a_l$ and $a_{l+1}$, including the small
lenses, see Figure 1.

The idea for
the next transformation  is to get rid of the jump across the real line.

We define $\tilde d$ as follows.
Let $\tilde d$ be analytic in $\Bbb C \setminus \cup J_l^-$, such that
the normal limits from above and below the real line
$d_{\pm} $ exist (apart from the endpoints) and such that
$$
\aligned
\tilde d_+ (\lambda) = \tilde d_-(\lambda) (1+|r(\lambda)|^2),~~~~~\lambda \in
\cup_l interior(J_l^-),
\\ lim_{\lambda \to \infty} d(\lambda) = 1.
\endaligned
$$
One can even give an explicit formula for $\tilde d$.
Denote $J_l^- = [a_l, b_l].$
Then
$$
\aligned
\tilde d(\lambda) =  exp~
[ \Sigma_l \int_{a_l }^{b_l} log  ( 1+|r(\zeta)|^2  )  {{d\zeta} \over
{2\pi i (\zeta-\lambda)}} ],~~~\lambda \in \Bbb C \setminus \cup J_l^- \endaligned
\tag21
$$
where the principal branch of the logarithm is chosen.

It is clear that the definitions above are not singular and indeed
$\tilde d, \tilde d^{-1}$ are bounded in the complex Riemann sphere.
In fact, it is easy to check that both $\tilde d(\lambda)$ and
$[\tilde d (\lambda^*)^*]^{-1}$ satisfy the conditions of the scalar
Riemann-Hilbert problem. Hence they must be equal, from which it
easily follows that
$$
\aligned
|\tilde d_-|^2 (1 + |r(\lambda)|^2) =1,~~~for~~ \lambda \in \cup_l J_l^-, \\
|\tilde d_-| \leq 1, ~~~for ~~~~\lambda \in \cup_l J_l^-,\\
|\tilde d_+| \leq 1 + sup_{\Bbb R} |r(\lambda)|^2 < \infty, ~~~for ~~~~\lambda \in \cup_l J_l
^-,
\endaligned
\tag21a
$$
and hence, by the maximum principle, $|\tilde d(\lambda)| \leq 
1 + sup_{\Bbb R} |r(\lambda)|^2 < \infty$. So 
$ |\tilde d(\lambda)|   $  is  bounded  uniformly  in the complex plane.
Similarly, $|\tilde d^{-1}(\lambda)|   $  is uniformly bounded  in the complex plane.

Near the points $a_l$ the local behavior of $\tilde d$  off the jump contour is
$$
\aligned
\tilde d \sim d_l (\lambda - a_l)^{ i \nu_l},
\endaligned
\tag21b
$$
and near $b_l$,
$$
\aligned
\tilde d \sim f_l (\lambda - b_l)^{ i \mu_l},
\endaligned
\tag21c
$$
where $\nu_l = { 1 \over {2\pi }} (1 + log | r(a_l) |^2), 
\mu_l = { 1 \over {2\pi }} (1 + log | r(b_l) |^2) $
and $d_l, f_l$ are independent of $\lambda$.
In fact
$$
\aligned
d_j=  exp~ [  \int_{-\infty }^{a_j} log (  z -a_j )  {{dlog ( 1+|r(z)|^2  )} \over {2 \pi i}}
+ \Sigma_{l \neq j} \int_{a_l }^{b_l} log  ( 1+|r(\zeta)|^2  )  {{d\zeta} \over
{2\pi i (\zeta-a_l)}} ], \\
f_j = 
 exp~ [ -\int_{-\infty }^{b_j} log (  z -b_j )   {{dlog ( 1+|r(z)|^2  )} \over {2 \pi i}}
+ \Sigma_ {l \neq j} \int_{a_l }^{b_l} log  ( 1+|r(\zeta)|^2  )  {{d\zeta} \over
{2\pi i (\zeta-b_l)}}],
\endaligned
\tag21d
$$
as can be shown by integration by parts.
The next transformation is then
$$
\aligned
Y(\lambda) = Z(\lambda) \Delta (\lambda),\\
where\\
\Delta (\lambda) =\pmatrix
&\tilde d(\lambda)   &0 \\
&0&  \tilde d^{-1} (\lambda)\endpmatrix.
\endaligned
\tag22
$$
Then $Y_+ = Y_- v_Y$, where
$$
\aligned
v_Y(\lambda)= \pmatrix
&e^{{g_+ - g_-}\over h}
&  -i~\tilde d^{-2} ~exp [{1 \over h} (\int log(\lambda -\eta) d\mu(\eta)
+g_-+g_- -2i\lambda x - 2i \lambda^2 t + X(\lambda))]\\
&0 &e^{{g_- - g_+}\over h}
\endpmatrix, \\
for~~~\lambda \in C,\\
v_Y(\lambda)= \pmatrix
&e^{{g_+ - g_-}\over h}
&0\\
&  i~\tilde d^2 ~exp [ {1 \over h} (-\int log(\lambda -\eta) d\mu(\eta)
-g_+-g_- +2i\lambda x + 2i \lambda^2 t + X^*(\lambda^*))]
&e^{{g_- - g_+}\over h} \endpmatrix,
\\for~~~~\lambda \in C^*,\\
v_Y(\lambda) = \pmatrix
&1  &-r(\lambda) \tilde d^2 e^{{2g -2i \lambda x - 2i\lambda^2 t} \over h} \\
&0&  1 \endpmatrix,~~~~for~~~~\lambda \in
\cup_l J_l^{+,up},~~~~~~~~~\\
v_Y(\lambda) = \pmatrix
&1  &0 \\
&r^* (\lambda^*) \tilde d^{-2} e^{{-2g + 2i \lambda x + 2i\lambda^2 t} \over h}&  1
\endpmatrix ,~~~~for~~~~\lambda \in
\cup_l J_l^{+,dn},\\
v_Y(\lambda) =
\pmatrix
&1   &0 \\
& \tilde d^{-2} {{r^* (\lambda^*)}  \over{ 1+|r(\lambda)|^2}}
e^{{-2g + 2i \lambda x + 2i\lambda^2 t} \over h}& 1
\endpmatrix,~~~ for~~~~\lambda \in
\cup_l J_l^{-,up},\\
v_Y(\lambda) =
\pmatrix
&1  &\tilde d^2 {{r(\lambda)} \over{ 1+|r(\lambda)|^2}}  e^{{2g -2i \lambda x - 2i\lambda^2 t
} \over h} \\
&0&  1 \endpmatrix,~~~ for~~~~\lambda \in  \cup_l J_l^{-,dn}. \\
\endaligned
\tag23
$$
Also $lim_{\lambda \to \infty} Y(\lambda) =I.$

At this point, we still have a  Riemann-Hilbert that is 
equivalent to the original one (11)-(12),
at least accepting the discrete-to-continuous passage in Theorem 1. 
Now, we can finally
start considering the limit  $h \to 0$.
Indeed, all terms involving $r(\lambda)$ and not supported on the real 
line can be neglected, not because
$r$ itself is small (it is not, see (16)-(16*)) but only
because it always appears 
multiplied by something exponentially small. 

In general jump matrices of the form $I + exponentially ~small$ 
can be neglected 
asymptotically as long as it is proved that the 
solution  is uniformly (in $x,t$) bounded as $h \to 0$.
We shall see  eventually that this is the case.

Assuming for the moment that this is true, 
we can simply delete the non-real part of the contour,
at least away from the endpoints of the intervals
$J_k$. Eventually (see section 5 later)
we can also delete the remaining small crosses centered
at such points.
We end up with a matrix valued function $W$, such that $W \sim Y$ near
infinity, and 
$$
\aligned
W_+(\lambda) = W_-(\lambda) v_W(\lambda),
\endaligned
\tag24
$$ 
where
$$
\aligned
v_W(\lambda)= \pmatrix
&e^{{g_+ - g_-}\over h}
&  -i~\tilde d^{-2} ~exp ({1 \over h} \int log(\lambda -\eta) d\mu(\eta))
exp (-{1 \over h} (-g_+-g_- +2i\lambda x + 2i \lambda^2 t - X(\lambda)))\\
&0 &e^{{g_- - g_+}\over h}
\endpmatrix, \\
for~~~\lambda \in C,\\
v_W(\lambda)= \pmatrix
&e^{{g_+ - g_-}\over h}
&0\\
&  i~\tilde d^2 ~exp ({-1 \over h} \int log(\lambda -\eta) d\mu(\eta))
exp ( {1 \over h} (-g_+-g_- +2i\lambda x + 2i \lambda^2 t + X^*(\lambda^*)))
&e^{{g_- - g_+}\over h} \endpmatrix,
\\for~~~~\lambda \in C^*.
\endaligned
\tag25
$$
Also $lim_{\lambda \to \infty} W(\lambda) =I.$ 

We will eventually check  that
the solution of the Riemann-Hilbert problem for $W$  exists and is
uniformly  (in $x,t$) bounded
as $h \to 0$.
This will justify neglecting the exponentially small terms in (23).
The passage from (23) to (25) 
will then be a posteriori justified, again at least except for the
remaining small crosses centered at a finite number of real points. 
Concerning the crosses see section 5 (Remark 5) and  Appendix A.1.

\bigskip

After applying the above transformation $Y \to W$, we  must  use another
lens transformation  to simplify the jump across the conotur
$C \cup C^*$. 
We will not describe this new lens transformation, 
since the discussion is exactly as in Chapter 4 of
[KMM]. We will simply state the end result of these  lens transformations
and the transformation  $Y \to W$.
The result is the so-called outer problem. We shall show in
the next section how it can be treated along the lines of Chapter 4 of [KMM].

\bigskip

THE OUTER PROBLEM.

We  first define the   analytic arcs 
$I_j, I^*_j,  j=1, ... , G/2$ as follows (they come in conjugate pairs).
Let the points  $\lambda_j,~~~j=0, ... , G$, in the open uper half-plane
be the branch points of the function $g$.
(The fact that there are $G+1$ of them is a consequence of
the definition of $g$, according to the "finite genus ansatz"; see below.) All such points lie
on the contour $C$ and we order them as
$\lambda_0, \lambda_1, ... , \lambda_G$,
according to the direction given to $C$.
The points $\lambda_0^*, \lambda_1^*, ... , \lambda_G^*$ are their
complex conjugates.
Then let
$I_0 = [0, \lambda_0]$ be the subarc of $C$ joining points $0$ and $\lambda_0$.
Similarly,
$ I_j = [\lambda _{2j-1}, \lambda _{2j}],~~~ j=1, ... , G/2$.
The points $\lambda_j,~~~j=0, ... , G$ lie in the open uper half-plane
and they are determined by a set of
transcendental equations that follow directly from the definitons of $g$ and $C$
(cf. Remark after Lemma 5.1.5 of [KMM]).
The connected components of the set $\Bbb C \setminus \cup_j (I_j \cup I^*_j)$
are the so-called "gaps", for example the gap $\Gamma_1$ joins $\lambda_0$
to $\lambda_1$, etc.  The subarcs $ I_j$ are  the "bands".

ASSUMPTION (G).  For simplicity, we will make the generic (in $x,t$) assumption that $\lambda_j \neq iA, ~j=0, ....., G.$

The finite genus ansatz implies that for each $x, t$ there is a finite positive integer
$G$ such that the contour $C$ can be divided into bands and gaps as above.
In fact, it follows from the conditions defining
$\rho, C$ that the measure reality condition
($\rho(\eta) d\eta~~~real$; see (18b))
splits into a measure strict negativity condition 
in the bands and a measure zero condition in the
gaps. Furthermore, the function $\theta (\lambda)$ 
of (18a) defined on $C$ is constant on each of the gaps
$\Gamma_j$, taking a value which we will denote by 
$\theta_j$, while the function $\phi$ of (18a) is constant
on each of the bands,
taking the value denoted by $\alpha_j$ on the band $I_j$.
For the justification of the finite ansatz 
under the barrier data see Appendix 2.

We are seeking a matrix $O$, which is analytic everywhere except across the contour 
$C\setminus \Gamma_{G/2 + 1}$ and its conjugate,
with  limits that are $L^2 (C\setminus
\Gamma_{G/2 + 1})$, 
converging to the identity at infinity and such that
$$
\aligned
O_+ (\lambda) = O_- (\lambda) \pmatrix &0 &i \tilde d^{-2} exp (-i \alpha_k/h) \\
& i \tilde d^2 exp (i \alpha_k /h) &0 \endpmatrix,~~~~\lambda \in I_k \cup I_k^*, \\ 
k=0, 1, ... , g/2,\\
O_+ (\lambda) = O_- (\lambda)  \pmatrix &exp (i\theta_k/h) &0 \\
&0 & exp (-i \theta_k/h) \endpmatrix , 
~~~~~~~~\lambda \in \Gamma_k \cup \Gamma_k^*,\\ k = 1, ... , G/2. 
\endaligned
\tag26
$$

We here recapitulate the sequence of matrix deformations
inroduced so far:

$$
\aligned
M (discrete) \to M(continuous) \to Q \to Z \to Y \to W \to O.
\endaligned
$$
The first  problem in the sequence, for $M$ (in its discrete version), is equivalent to the
inverse scattering problem for NLS.
The last Riemann-Hilbert problem, for $O$, (26), will be  solved explicitly via theta
functions (see (33) of section 5).

\bigskip

4. THE RESULT

\bigskip

As explained in Appendix 2,
the finite genus ansatz holds for the semiclassical asymptotics
under barrier initial data, at least generically.

Assuming the finite genus ansatz,
the $x,t$-plane can be divided into (possibly empty) open
regions $R_G, G=0,2,4,.....$ (here $G/2+1$ is the number of 
components of the support of the equilibrium measure of $C$, see
Appendix A.2),
together with their
boundaries, such that within each region
the asymptotics of the solution of (1)
with barrier data  as in (2) can be given as follows.

\bigskip

THEOREM 2.
Let $x_0, t_0$ lie in region $R_G$.
The solution $u^{h}(x,t)$  is asymptotically
described  (locally) as a slowly
modulated $G+1$ phase wavetrain.  Setting $x=x_0+h \hat{x}$ 
and $t=t_0+h \hat{t}$,
so that $x_0, t_0$ are "slow" variables
while $\hat{x}, \hat{t}$ are "fast" variables,
there exist
parameters

$a,  U = (U_0, U_1, .... , U_G)^T,$
$k =(k_0, k_1, ......, k_G)^T,$
$w =(w_0, w_1, ....., w_G)^T, $
$Y =(Y_0, Y_1, ........., Y_G)^T,$
$Z =( Z_0, Z_1, ...... , Z_G)^T $
depending on the slow variables
$x_0$ and $t_0$  and possibly $h$ (but not  $\hat{x}, \hat{t}$)
such that
$$
\aligned
u^h (x,t)= u^h (x_0 + h \hat{x},  t_0 + h \hat{t}) \sim
a(x_0, t_0) e^{iU_0(x_0, t_0)/h}
e^{i(k_0(x_0, t_0) \hat{x}-w_0(x_0, t_0) \hat{t})} \\
\cdot \tilde d^2(\lambda_0) \frac{\Theta(  Y(x_0, t_0)+
i  U(x_0, t_0)/h +
i(  k(x_0, t_0) \hat{x}-  w(x_0, t_0)\hat{t}))}
{
\Theta(  Z(x_0, t_0)+
i  U(x_0, t_0)/h +
i(  k(x_0, t_0)\hat{x}-  w(x_0, t_0)\hat{t}))}.\endaligned
\tag27
$$
Unlike the analogous formula in [KMM], we allow here a dependence of the
parameters on $h$. But of course, we can always rearrange terms to arrive at 
a formula like (27) where the parameters are independent of $h$.

All parameters can be defined in terms of an underlying
Riemann surface $X$.
The moduli of $X$ are given by $\lambda_j, ~ j=0, .... , G$ and their complex
conjugates  $\lambda_j^*, ~ j=0, .... , G$.
The genus of $X$ is $G$. The moduli of $X$ vary slowly with $x,t$, i.e.
they depend on  $x_0, t_0$ but not
$\hat{x}, \hat{t}$. For the exact formulae 
for the parameters
as well as the definition  of the theta functions we present the following construction.
 
The Riemann surface $X$ is constructed by cutting two copies of the complex
sphere along the slits
$I_0 \cup I_0^*, I_j, I_j^*, j=1. ... ,G$,
and pasting the "top" copy to the "bottom" copy along these
very slits.

In exact analogy with the discussion of  [KMM] (as  exemplified
in Figure 4.5 on p.62 of [KMM]) we define the homology cycles
$a_j, b_j, ~~~ j=1, ... , G$
as follows.
Cycle $a_1$ goes around
the slit $I_0 \cup I_0^*$ joining $\lambda_0$ to
$\lambda_0^*$,
remaining on the top sheet, oriented counterclockwise,
$a_2$ goes through  the slits $I_{-1}$ and $I_1$
starting from the top sheet, also
oriented counterclockwise,
$a_3$ goes around the slits $I_{-1}, I_0 \cup I_0^*, I_1$
remaining on the top sheet, oriented counterclockwise, etc.
Cycle $b_1$ goes through $I_0$  and $I_1$ oriented counterclockwise,
cycle  $b_2$ goes through $I_{-1}$  and $I_1$, also
oriented counterclockwise,
cycle  $b_3$ goes through $I_{-1}$  and $I_2$, and
around the slits $I_{-1}, I_0 \cup I_0^*, I_1$,
oriented counterclockwise, etc. 
\bigskip

Let 
$$
\aligned
R(\lambda)^2 = \prod_{k=0}^{G}(\lambda-\lambda_k)(\lambda-\lambda_k^*),
\endaligned
$$
choosing the particular branch that is cut along the bands $I_k$ and
$I_k^*$ and such that
$$
\aligned
\lim_{\lambda\rightarrow\infty}\frac{R(\lambda)}{\lambda^{G+1}}= -1.
\endaligned
$$
On $X$
there is a complex $G$-dimensional linear space of holomorphic
differentials, with basis elements $\nu_k(P)$ for $k=1,\dots,G$ that
can be written in the form
$$
\aligned
\nu_k(P)=\frac{\displaystyle\sum_{j=0}^{G-1}c_{kj}\lambda(P)^j}
{R_X(P)}\,d\lambda(P)\,,
\endaligned
$$
where $R_X(P)$ is a ``lifting'' of the function $R(\lambda)$ 
from the cut plane to $X$: if $P$ is on the first sheet of $X$ then
$R_X(P)=R(\lambda(P))$ and if $P$ is on the second sheet of
$X$ then $R_X(P)=-R(\lambda(P))$.  The coefficients $c_{kj}$
are uniquely determined by the constraint that the differentials
satisfy the normalization conditions:
$$
\aligned
\oint_{a_j}\nu_k(P) = 2\pi i\delta_{jk}.
\endaligned
$$
From the normalized differentials, one defines a $G\times G$ matrix
$H$ (the period matrix) by the formula:
$$
\aligned
H_{jk}=\oint_{b_j}\nu_k(P).
\endaligned
$$
It is a consequence of the standard theory of Riemann surfaces that
$ H$ is a symmetric matrix whose real part is  negative
definite.
 
In particular, we can
define the theta function
$$
\aligned
\Theta( w):=\sum_{  n\in {\Bbb Z}^G}\exp({1 \over 2} n^T  H
n+ n^T  w),
\endaligned
$$
where $ H$ is the period matrix associated to $X$.
Since the real part of $ H$ is  negative
definite, the series converges.
 
We arbitrarily fix a base point $P_0$ on $X$.  The
Abel map
$ A:X\to Jac(X)$ is then defined componentwise as follows:
$$
\aligned
A_k(P;P_0):=\int_{P_0}^P\nu_k(P'),~~~~ k=1,\dots,G,
\endaligned
$$
where $P'$ is an integration variable.
 
\bigskip

A particularly important element of the Jacobian is the
Riemann constant vector
$K$ which is defined, modulo the lattice
$\Lambda$, componentwise by
$$
\aligned
K_k:=\pi i + \frac{H_{kk}}{2}-\frac{1}{2\pi i}\sum_{j=1\atop j\neq
k}^G\oint_{a_j} \left(\nu_j(P)\int_{P_0}^P\nu_k(P')\right),
\endaligned
$$
where the index $k$ varies between $1$ and $G$.

Next, we will need to define a certain meromorphic differential
on $X$.  Let $\Omega(P)$ be holomorphic away from the points $\infty_1$
and $\infty_2$, where it has the behavior
$$
\aligned
\Omega(P) = dp(\lambda(P)) + \displaystyle
\left(\frac{d\lambda(P)}{\lambda(P)^2}\right), ~~~P\to
\infty_1,\\
\Omega(P) =-dp(\lambda(P))+\displaystyle
O\left( \frac{d\lambda(P)}{\lambda(P)^2}\right), ~~~~
P\to \infty_2,
\endaligned
$$
and made unique by the normalization conditions
$$
\aligned
\oint_{a_j}\Omega(P) = 0, j=1,\dots,G.
\endaligned
$$
Here $p$ is some polynomial. In the present context see section 5, equation (31) for its
definition.

Let the vector $U\in {\Bbb C}^G$ be defined
componentwise by
$$
\aligned
U_j:=\oint_{b_j}\Omega(P).
\endaligned
$$
Note that $\Omega(P)$ has no residues.

Let the  vectors $ V_1,  V_2$ be defined componentwise by
$$
\aligned
V_{1,k}=
(A_k(\lambda_{1+}^*)+A_k(\lambda_{2+})+A_k
(\lambda_{3+}^*)+\dots+A_k(\lambda_{G+}))
+A_k(\infty) +\pi i + \frac{H_{kk}}{2},\\
V_{2,k}=(A_k(\lambda_{1+}^*)+A_k(\lambda_{2+})+A_k
(\lambda_{3+}^*)+\dots+A_k(\lambda_{G+}))
-A_k(\infty) +\pi i + \frac{H_{kk}}{2},
\endaligned
$$
where $k =1, ..., G$, and  the $+$ index means that the integral for $A$
is to be taken on the first sheet of $X$, with base point $\lambda_0.$

Finally, let
$$
\aligned
a=\frac{\Theta( Z)}
{\Theta(  Y)}
\sum_{k=0}^G(-1)^k\Im(\lambda_k)  ,\\
k_n=\partial_x U_n,~~~~~
w_n=-\partial_t U_n,~~~~~
n=0,\dots,G,
\endaligned
$$
where
$$
\aligned 
Y=- A(\infty)-  V_1,~~~~ 
Z=  A(\infty)- V_1,
\endaligned
$$
and  $ U_0 =-(\theta_1 + \alpha_0)$
where $\theta_1$ is the (constant in $\lambda$) value of the function $\theta$ in the gap $\Gamma_1$
and $\alpha_0$ is the (constant) value of the function $\phi$ in the band $I_0$. 
The fact that these values 
are actually constants in $\lambda$ follows from the conditions defining $g$ and $C$.
 
Now, the parameters appearing in formula (27) are completely described.
 
We simply note  here that the $U_i$ and hence the $k_i$ and $w_i$ are real
modulo $O(h)$.  We also note that the denominator in (27) never vanishes (for any
$x_0, t_0, \hat{x}, \hat{t}$).
 
\bigskip

REMARK: Because $C$ depends only on the eigenvalue density $\rho^0$ and not on
the reflection coefficient $r$, the constructions of $C$, the Riemann surface $X$,
the holomorphic differential $\nu_k(P)$, the Abel map $A$ and the theta functions
are all independent of $r$. The only contribution of $r$  
comes through the factors $\tilde d^2, \tilde d^{-2}$ in (26) and the factor
$\tilde d^2 (\lambda_0)$ in (27). This is why our discussion in section 4 is virtually 
repeating verbatim  the analogous discussion of section 4 in [KMM].

\bigskip

5. REMARKS AND PROOF OF THEOREM 2.

\bigskip

1. Formula (27) is locally a so-called finite gap expression. It describes 
violent oscillations of bounded amplitude but high frequency $O(1/h)$.

2. As in [KMM] weak limits of densities exist:
$\rho= lim_{h \to 0} |u^{h}|^2$ and
$\mu =   lim_{h \to 0} {-i h \over 2} (\bar u^{h}  u^{h}_x -
 u^{h}  \bar u^{h}_x )$.
These limits are actually strong in the genus zero region.

3. Naturally, since the initial data is discontinuous and the limiting
Euler system (1a) is elliptic (see [KMM]) one expects the break-time of
the limiting system to be zero. This is indeed the case, as numerical
experiments by H.Ceniceros and F.Tian have shown [T],
or as can be shown analytically by considering the limiting 
Euler system directly:  from the second equation of (1a) it is obvious
that $\mu_t$ is infinite at
$x=\pm{1 \over 2}, t=0$. Of course a genus zero
region still exists but the first caustic 
(the boundary between the genus zero region and higher genus regions) 
touches the $t=0$ axis of the $x,t$-plane at $x = \pm 1/2$.

4. The proof of the results in [KMM] makes use of the assumption that the 
eigenvalue density can be analytically extended in the upper half-plane
with the spike where eigenvalues accumulate deleted.
This $is$ the case here, see formula (9a). 
The branch root singularity at $iA$ 
is integrable. Integrals like $\int \rho^0 d\eta$ or
$\int log(\lambda-\eta) \rho^0 (\eta) d\eta$ can still be deformed
and the Cauchy theorem holds. On the other hand, we assume that $x,t$ are such
that $\lambda_j \neq iA, ~j=0,1,....g, $ a generic case.

5. A priori, neglecting the exponentially terms in (23) 
only allows us to "delete" arcs $J_l^{+,up},
J_l^{+,dn}, J_l^{-,up}, J_l^{-,dn}$ only away from the real endpoints 
of $J_l^+, J_l^-$. For entirely rigorous justifications of the deletion of the small crosses 
centered at each such point remaining after the deletion of the bulk of  
the  arcs $J_l^{+,up},
J_l^{+,dn}, J_l^{-,up}, J_l^{-,dn}$, one must construct local parametrices of the
Riemann-Hilbert problem and make sure they match with the solution of (25)
away from the endpoints.  
This is a procedure that is by now standard in the literature;
more details are given in the Appendix. In fact, essentially the same
situation has appeared in [DZ1]. The "local" Riemann-Hilbert problem can be solved
via parabolic cylinder functions. (In [DZ1], of course, the contribution was not negligible,
since one was trying to evaluate the term of order $O(t^{-1/2})$ of the asymptotics.
Also in [DZ1] the reflection coefficient is independent of
the parameter  $1/t$ going to zero,  but this is irrelevant since the
local Riemann-Hilbert problem is solved exactly via parabolic cylinder functions.)

6. As in [KMM], one needs to provide a
local parametrix near the origin  and then match it
with the solution of the "outer problem". This can still be done, 
using the Fredholm theory described in [KMM].  The "cyclic" relation,
that the product of limits of jumps at $0$ is the identity, still holds.
In view of (9b) the cyclic relation follows easily from the
analogous relation in the reflectionless case. The discussion in
sections (4.4.3), (4.5.1) and
(4.5.2) of  [KMM] can be then followed verbatim.

7. The Bohr-Sommerfeld condition (8) is not quite the same
as the condition postulated in [KMM] defining the so-called soliton ensemble,
which  translates as
$$
\aligned
(A^2+\lambda_k^2)^{1/2} = h k \pi
\endaligned
$$
for our present problem (1)-(2). 
Now, one can notice that the difference between the
two conditions gives rise to a uniform error of
higher order $O(h^2)$.
The analysis in [KMM] (Chapter 3) of estimates needed for the 
passage from a "discrete" Riemann-Hilbert problem to a 
"continuum"  Riemann-Hilbert problem
is not altered by this innocent modification.

8. From  formulae (6)-(7) it is obvious that $\alpha(0) =0$ if
$A h $ is an odd multiple of $\pi/2$. So, for a particular sequence of $h$ going to zero
we have a $spectral~singularity$ at the real point $0$. The Riemann-Hilbert problem jump
becomes singular at $0$. By assumption we have excluded such values of
$h$. We plan to study 
the effect of  a real spectral singularity on the semiclassical
behavior of the focusing NLS in a later publication.

9. The proof of Theorem 2 now follows the discussion of Chapter 4 of [KMM].
One minor change is the extra factor $\tilde d^{\pm 2}$ appearing in the
off-diagonal terms of (26). This factor can be taken care of by the auxiliary
scalar Riemann-Hilbert problem (4.38) in section 4.3.1 of [KMM].
The proof goes through with only a minor change:
$\alpha_k $ has to be substituted by
$\alpha_k +2i h ~log \tilde d$. Of course $\tilde d$ 
has to appear in the final formula (27).
Granted that the term $h \tilde d$ is no more constant on bands, but 
then the auxiliary scalar
Riemann-Hilbert problem is still explicitly solvable. 

More precisely, we introduce the scalar problem:
$$
\aligned
H_+(\lambda)-H_-(\lambda)=- \theta_k,~~~
\lambda\in\Gamma_k\cup\Gamma_{k}^*,~~~~~&k=1,\dots,G/2\,,\\
H_+(\lambda)+H_-(\lambda)=-(\alpha_k + 2i h log \tilde d),~~~~~~
\lambda\in I_k\cup I_{k}^*,~~~~~~&k=0,\dots,G/2\,, 
\endaligned
\tag 28
$$
not specifying any condition at infinity yet.
Consider the matrix defined by
$$
\aligned
P(\lambda)= O(\lambda)\exp(iH(\lambda)\sigma_3/h), 
\endaligned
$$
where $O$ is the solution of the outer Riemann-Hilbert problem 
(26).
It is straightforward to verify that the matrix $ P(\lambda)$ 
has the identity matrix as the jump matrix in all
gaps $\Gamma_k$ and $\Gamma_k^*$.  Since the boundary values of
$ O(\lambda)$ and $H(\lambda)$ are 
continuous, it follows that $ P(\lambda)$ is in fact
analytic in the gaps.  In the bands, the jump relation becomes simply
$$
\aligned
P_+(\lambda)=i P_-(\lambda)
\sigma_1,
\endaligned
$$
so the jump relation is the same in all bands.  Next, suppose that
$\beta(\lambda)$ is a scalar function analytic in the
$\lambda$-plane except at the bands, where it satisfies
$
\beta_+(\lambda)=-i\beta_-(\lambda).  $
Suppose further for the sake of concreteness that
$\beta(\lambda)\rightarrow 1$ as $\lambda\rightarrow\infty$.
Then, setting
$$
\aligned
V(\lambda)=\beta(\lambda) P(\lambda),
\endaligned
\tag29
$$
we see that the jump relations for $ V(\lambda)$ take on the
elementary form:
$$
\aligned
V_+(\lambda)= V_-(\lambda)\sigma_1,~~~~
\lambda\in \cup_k (I_k \cup I_k^*).
\endaligned
\tag30
$$
Our purpose in reducing the jump relations to this universal constant
form is that it can be explicitly solved in terms of theta functions.

But let us describe the
scalar functions $H(\lambda)$ and $\beta(\lambda)$. We get
$$
\aligned
\beta(\lambda)^4=\frac{\lambda-\lambda_0^*}{\lambda-\lambda_0}
\prod_{k=1}^{G/2}\frac{\lambda-\lambda_{2k-1}}{\lambda-\lambda^*_{2k-1}}
\cdot\frac{\lambda-\lambda^*_{2k}}{\lambda-\lambda_{2k}}\,,
\endaligned
$$
and for $\beta(\lambda)$ we select the branch that tends to
unity for large $\lambda$ and that is cut along the bands $I_k$ and
$I_k^*$.  It is easily checked that $\beta(\lambda)$ as defined
here is the only function satisfying the required jump condition and
normalization at infinity that has continuous boundary values (except
at half of the endpoints).  
To find $H(\lambda)$, we introduce the function $R(\lambda)$
defined by
$$
\aligned
R(\lambda)^2 = \prod_{k=0}^{G}(\lambda-\lambda_k)(\lambda-\lambda_k^*),
\endaligned
$$
choosing the particular branch that is cut along the bands $I_k$ and
$I_k^*$ and satisfies
$$
\aligned
\lim_{\lambda\rightarrow\infty}\frac{R(\lambda)}{\lambda^{G+1}}= -1, 
\endaligned
$$
This defines a real function,  i.e. one that satisfies
$R(\lambda^*)=R(\lambda)^*$.  At the bands, we have
$R_+(\lambda)=-R_-(\lambda)$, while $R(\lambda)$ is analytic in the
gaps.  Set
$$
\aligned
H(\lambda)=k(\lambda)R(\lambda),
\endaligned
$$
where
$$
\aligned
k(\lambda)=\frac{1}{2\pi i}
\sum_{n=1}^{G/2}\theta_n\int_{\Gamma_n\cup\Gamma_n^*}
\frac{d\eta}{(\lambda-\eta)R(\eta)} +
\frac{1}{2\pi i}\sum_{n=0}^{G/2}
\int_{I_n\cup I_n^*}\frac{(\alpha_n + 2ih log \tilde d (\eta))
~~d\eta}{(\lambda-\eta)R_+(\eta)}.
\endaligned
$$
We see that $k(\lambda)$ satisfies the jump relations:
$$
\aligned
k_+(\lambda)-k_-(\lambda)= 
\displaystyle -\frac{\theta_n}{R(\lambda)},~~~~\lambda\in
\Gamma_n \cup\Gamma_n^*\\
k_+(\lambda)-k_-(\lambda)=
\displaystyle -\frac{\alpha_n + 2i h log \tilde d(\lambda)}{R_+(\lambda)},~~~~\lambda\in
I_n\cup I_n^*,
\endaligned
$$
and is otherwise analytic. 
So $H$ satisfies (28).

The function $k$ blows up like $(\lambda-\lambda_n)^{-1/2}$ near each
endpoint, has continuous boundary values in between the endpoints, and
vanishes like $1/\lambda$ for large $\lambda$.  It is the only such
solution of the jump relations. 
The factor of $R(\lambda)$ renormalizes
the singularities at the endpoints, so that, as desired, the boundary
values of $H(\lambda)$ are bounded continuous functions.  Near
infinity, there is the asymptotic expansion:
$$
\aligned
H(\lambda)=H_G\lambda^G + H_{G-1}\lambda^{G-1} +
\dots + H_1\lambda + H_0 + O (\lambda^{-1})\\
=p(\lambda) + O (\lambda^{-1}),
\endaligned
\tag31
$$
where all coefficients $H_j$ of the polynomial $p(\lambda)$
can be found explicitly by expanding $R(\lambda)$ and the Cauchy
integral $k(\lambda)$ for large $\lambda$.  It is easy
to see from the reality of $\theta_j$ and $\alpha_j$ that
$p(\lambda)$ is a polynomial with  coefficients which are real modulo
$O(h)$.

So the matrix function $V$ defined in (29) has the following asymptotics at infinity:
$$
\aligned
V(\lambda) exp[-ip(\lambda) \sigma_3/h] = I + O(\lambda^{-1}).
\endaligned
\tag 32
$$
This together with the jump relations (30) 
defines a Riemann-Hilbert problem for $V$ that
can be explicitly solved via theta functions. 
Equivalently $P$ and $O$ can be explicitly expressed in terms of
theta functions. For example,
it is now elementary to check that
the solution of the outer  Riemann-Hilbert
problem (26) for $O$  is given by the formulae:
$$
\aligned
O_{11}(\lambda)=\displaystyle
\frac{b^-(\lambda)}{\beta(\lambda)}
\frac{\Theta( A(\infty)- V_1)}
{\Theta( A(\lambda)- V_1)}
\frac{\Theta(A(\lambda)- V_1+iU/\hbar)}
{\Theta( A(\infty)- V_1+i U/\hbar)},\\
O_{12}(\lambda)=\displaystyle
\frac{b^+(\lambda)}{\beta(\lambda)}
e^{2iR(\lambda)k_+(\lambda_0)/\hbar}
\frac{\Theta( A(\infty)- V_1)}
{\Theta(- A(\lambda)- V_1)}
\frac{\Theta(- A(\lambda)- V_1+i U/\hbar)}
{\Theta(A(\infty)- V_1+i U/\hbar)},\\
O_{21}(\lambda)=\displaystyle
\frac{b^+(\lambda)}{\beta(\lambda)}
e^{-2iR(\lambda)k_+(\lambda_0)/\hbar}
\frac{\Theta(- A(\infty)- V_2)}
{\Theta( A(\lambda)- V_2)}
\frac{\Theta(f A(\lambda)- V_2+i U/\hbar)}
{\Theta(- A(\infty)- V_2+i U/\hbar)},\\
O_{22}(\lambda)=\displaystyle
\frac{b^-(\lambda)}{\beta(\lambda)}
\frac{\Theta(- A(\infty)- V_2)}
{\Theta(- A(\lambda)- V_2)}
\frac{\Theta(- A(\lambda)- V_2+i U/\hbar)}
{\Theta(-A(\infty)- V_2+i U/\hbar)},
\endaligned
\tag33
$$
where
$$
\aligned
b^\pm(\lambda)=\frac{R(\lambda)\pm (\lambda-\lambda_0^*)(\lambda-\lambda_1)\dots 
(\lambda-\lambda_G^*)} {2R(\lambda)}.
\endaligned
\tag34
$$
From the explicit solution,
using formula (14) with the obvious substitution of $O$ for $M$ one derives
formula (27). This completes the proof of Theorem 2.

We end this section by once more recapitulating
the sequence of matrix deformations used:
$$
\aligned
M (discrete) \to M(continuous) \to Q \to Z \to Y \to W \to O . 
\endaligned
\tag35
$$
The first  problem in the sequence, for $M$ (in its discrete version), is equivalent to the
inverse scattering problem for NLS.
The last Riemann-Hilbert problem, for $O$, (26), was  solved explicitly via theta
functions.
Since, as is seen, the solution  $O$  of the outer problem 
is uniformly bounded in $x,t, $ with $L^2$ limits $O_+, O_-$, as $h \to 0$, so are
$W$ and $Y$  and  the (asymptotically valid) transformations from $Y$ to $O$ and back 
are a posteriori justified.

\newpage

APPENDIX 1. THE CROSS PROBLEM.

\bigskip

The transformation from the Riemann-Hilbert problem (23) to the  problem (24)-(25)
requires two steps. First, the deletion of the lens contours away from the points
$a_l, b_l$. This is immediate because the jump matrices are uniformly exponentially
small perturbations of the identity.

Second, one needs to consider the small remaining crosses centered at the points $a_l, b_l$
(see the remark of section 5). Since the jump matrices are not uniformly small there, one needs to
find "local" parametrices. In other words, one needs to solve
the local Riemann-Hilbert problems. 

For example, after translation 
all  problems centered at $a_l$
look as follows:
to find a matrix $L$ which is analytic in $\Bbb C \setminus \Gamma$ where $\Gamma$ is the cross
shown in Figure A.1,  centered at $0$. The actual angles between the four half-lines
emanating from 0 are not important as long as every half-line is in a different quadrant. 

\bigskip

\epsfxsize=10truecm
\epsffile{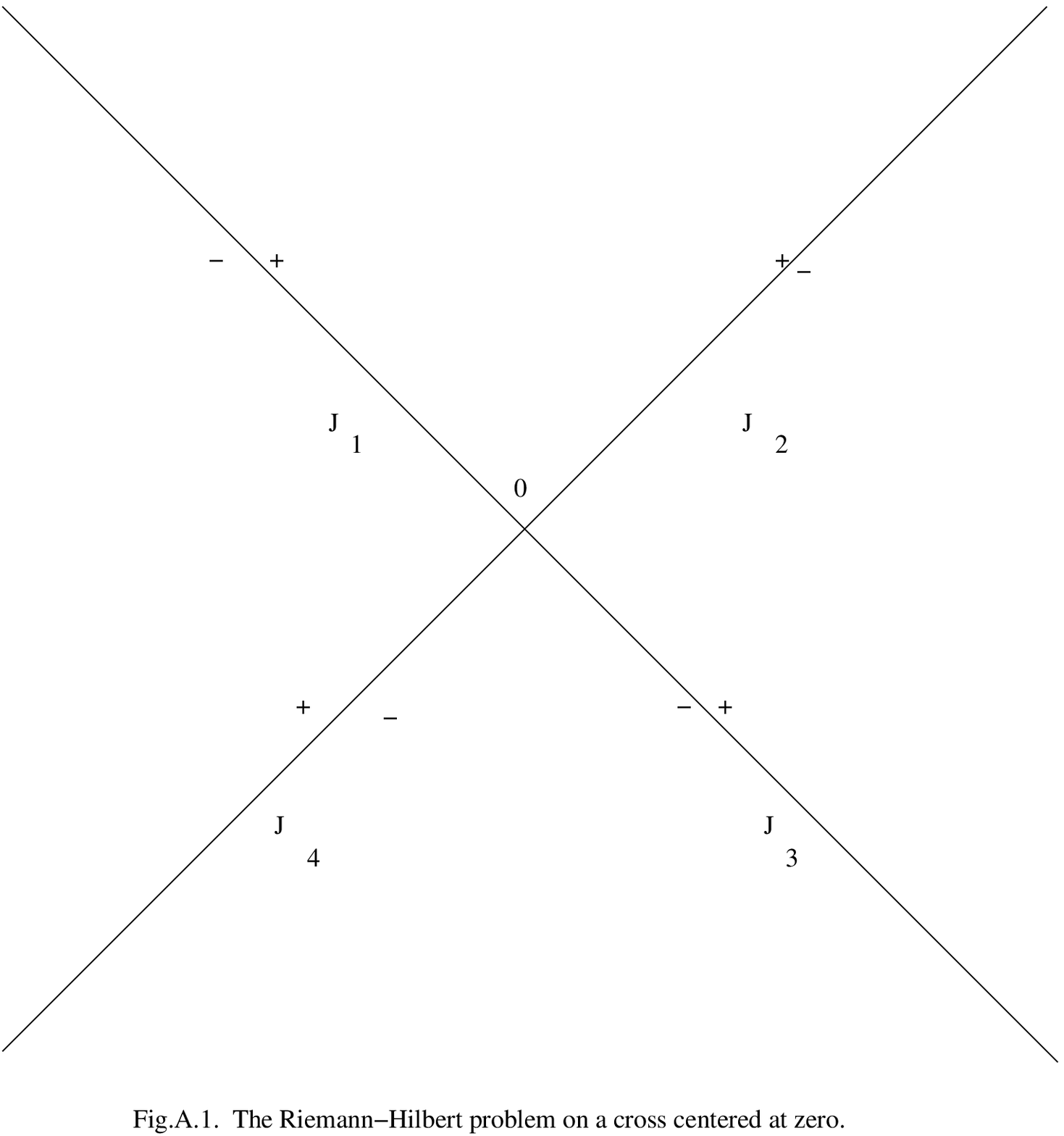}

\newpage

The jumps for $L$ are

$$
\aligned
L_+=L_- v_L(\lambda),~~~where\\
v_L(\lambda) =
\pmatrix
&1
&d_l^2 {r_l \over{ 1+|r_l|^2}} \lambda^{2\nu_l i} e^{{2g_l -2i \lambda x - 2i\lambda^2 t} \over h} \\
&0&  1 \endpmatrix,~~~ for~~~~\lambda \in   J_1,\\
v_L(\lambda) =
\pmatrix
&1   &0 \\
& d_l^{-2} {r^*_l  \over{ 1+|r_l|^2}} \lambda^{-2\nu_l i}
e^{{-2g_l + 2i \lambda x + 2i\lambda^2 t} \over h}& 1
\endpmatrix,~~~ for~~~~\lambda \in
J_2,\\
v_L(\lambda)= \pmatrix
&1  &- d_l^2 r_l \lambda^{2\nu_l i}  e^{{2g_l -2i \lambda x - 2i\lambda^2 t} \over h} \\
&0&  1 \endpmatrix,~~~~for~~~~\lambda \in
J_3,~~~~~~~~~\\
v_L(\lambda) = \pmatrix
&1  &0 \\
&  d^{-2}_l r_l^* \lambda^{-2\nu_l i}  e^{{-2g_l + 2i \lambda x + 2i\lambda^2 t} \over h}&  1
\endpmatrix ,~~~~for~~~~\lambda \in
J_4.\\
\endaligned
\tag A.1
$$
Also $lim_{\lambda \to \infty} L(\lambda) =I.$

Here
$$
\aligned
\nu_l = { 1 \over {2\pi }}  log (1+| r(a_l) |^2),\\
g_l = g(\lambda=a_l), r_l = r(a_l)
\endaligned
$$
and $d_l$ is defined by the local behavior of $\tilde d$ near $a_l$ (see (21b)).

After  a conjugation of the jump matrix by
$diag (d_l e^{{g_l +i{{x^2}\over {4t}}} \over h},
d_l^{-1} e^{{-g_l -i{{x^2}\over {4t}}} \over h})$,
a further translation $\lambda \to \lambda -x/2$ 
(to complete the square) and a rescaling
$\lambda \to \lambda  ({t \over h})^{1/2},$
we end up with a problem with jumps on a
rescaled cross which is no more small. Extending the cross to infinity (by setting
the jump equal to the identity on the extension) we have a new Riemann-Hilbert problem
which can be approximated by the following:
$$
\aligned
\Psi_+=\Psi_- v_{\Psi}, where ~~v_{\Psi}(\lambda)=\\
\pmatrix
&1
&{r_l  \over{ 1+|r_l|^2}}  \lambda^{2\nu_l i} e^{ - 2i\lambda^2 }  \\
&0&  1 \endpmatrix,~~~ for~~~~\lambda \in  \tilde J_1, \\
\pmatrix
&1   &0 \\
& {r^*_l  \over{ 1+|r_l|^2}}\lambda^{-2\nu_l i} 
e^{ 2i\lambda^2 } & 1
\endpmatrix,~~~ for~~~~\lambda \in \tilde
J_2,\\
\pmatrix
&1  &- r_l \lambda^{2\nu_l i}  e^{ - 2i\lambda^2 } \\
&0&  1 \endpmatrix,~~~~for~~~~\lambda \in \tilde
J_3,~~~~~~~~~\\
\pmatrix
&1  &0 \\
&r^*_l \lambda^{-2\nu_l i}  e^{ 2i\lambda^2 }&  1
\endpmatrix ,~~~~for~~~~\lambda \in \tilde
J_4,\\
lim_{\lambda \to \infty} S(\lambda) = I,
\endaligned
\tag A.2
$$
where $\tilde J_i$ is the extension of  $J_i,~~ i=1,2,3,4.$

The last Riemann-Hilbert problem can be explicilty solved via parabolic cylinder
functions (see e.g. [DZ1]).
Indeed, let $D_a(\lambda)$ denote the standard parabolic cylinder function. 
Then $D_a(e^{-3i \pi /4}\lambda)$ and $D_a(-e^{-3i \pi /4}\lambda)$ solve the ODE
$$
\aligned
{{d^2 D} \over {d \lambda^2}} +({1 \over 2} -{\lambda^2 \over 4} + a) D =0,
\endaligned
$$
where $a = i \nu_l$.
Let
$$
\aligned
S=
\pmatrix
&S_{11} &S_{12} \\
&S_{21} &S_{22}\endpmatrix,
\endaligned
\tag A.3
$$
where
$$
\aligned
S_{11}= 
e^{-3\pi \nu_l/4} D_a (e^{-3i \pi /4}\lambda),\\
S_{12}= {{r_l^* \Gamma (a)}   \over { (2\pi)^{1/2} e^{-i \pi/4}}} e^{3\pi \nu_l/4}
[{d \over {d\lambda}} D_{-a} (e^{-i \pi /4}\lambda) -{{i\lambda }\over {2}} 
D_{-a} (e^{-i \pi /4} \lambda) ],\\
S_{21}= {{r_l \Gamma (-a)}   \over{ (2\pi)^{1/2} e^{i \pi/4}}} e^{-\pi \nu_l/4} 
[{d \over {d\lambda}} D_a(e^{-3i \pi /4}\lambda) + {{i\lambda \over 2}} D_a(e^{-3i \pi /4}\lambda) ] ,\\ 
S_{22}=  e^{\pi \nu_l/4} D_a(e^{-i \pi /4}\lambda) ,\\ 
for~~~~Im \lambda >0,
\endaligned
\tag A.4
$$
and
$$
\aligned
S_{11}=
e^{\pi \nu_l/4} D_a (e^{i \pi /4}\lambda),\\
S_{12}= {{r_l^* \Gamma (a)}   \over{ (2\pi)^{1/2} e^{-i \pi/4}}}e^{-\pi \nu_l/4}
[{d \over {d\lambda}} D_{-a}(e^{3i \pi /4}\lambda) -{{i\lambda \over 2}}D_{-a}(e^{3i \pi /4}\lambda) ],\\
S_{21}= {{r_l \Gamma (-a)}   \over{ (2\pi)^{1/2} e^{i \pi/4}}} e^{3\pi \nu_l/2}
[{d \over{d\lambda}} D_a(e^{i \pi /4}\lambda) +{{i\lambda \over 2}}D_a(e^{i \pi /4}\lambda) ] ,\\ 
S_{22}=  e^{-3\pi \nu_l/4} D_a(e^{3i \pi /4}\lambda) ,\\ 
for~~~~Im \lambda <0.
\endaligned
\tag A.5
$$
Then it is possible to check that
$$
\aligned
S_+=S_- v_{S}, where ~~S(\lambda)=\\
\pmatrix
&1
&{r_l  \over{ 1+|r_l|^2}}  \\
&0&  1 \endpmatrix,~~~ for~~~~\lambda \in  \tilde J_1, \\
\pmatrix
&1   &0 \\
& {r^*_l  \over{ 1+|r_l|^2}} & 1
\endpmatrix,~~~ for~~~~\lambda \in \tilde
J_2,\\
\pmatrix
&1  &- r_l  \\
&0&  1 \endpmatrix,~~~~for~~~~\lambda \in \tilde
J_3,~~~~~~~~~\\
\pmatrix
&1  &0 \\
&r^*_l &  1
\endpmatrix ,~~~~for~~~~\lambda \in \tilde
J_4,\\
lim_{\lambda \to \infty} S(\lambda) = I.
\endaligned
\tag A.6
$$
Now, setting
$$
\aligned
S = \Psi \lambda^{i \nu_l \sigma_3} e^{-i \lambda^2/4~\sigma_3}.
\endaligned
\tag A.7
$$ 
and using the well known asymptotics for the parabolic cylinder function at infinity 
$$
\aligned
D_a(\lambda)= \lambda^a e^{- \lambda^2/4} (1 +O(\lambda^{-2})),~~~|arg\lambda|< {{3\pi}\over 4},\\
D_a(\lambda)= \lambda^a e^{- \lambda^2/4} (1 +O(\lambda^{-2}))-
{{(2\pi)^{1/2} } \over {\Gamma (-a) }} e^{i\pi a} \lambda^{-a-1} e^{ \lambda^2/4}(1 +O(\lambda^{-2})),~~~
{{\pi}\over 4} <arg\lambda< {{5\pi}\over 4},\\
D_a(\lambda)= \lambda^a e^{- \lambda^2/4} (1 +O(\lambda^{-2}))-
{{(2\pi)^{1/2} } \over {\Gamma (-a) }} e^{-i\pi a} \lambda^{-a-1} e^{ \lambda^2/4}(1 +O(\lambda^{-2})),~~~
{{-5\pi}\over 4} <arg\lambda< {{-\pi}\over 4},
\endaligned
\tag A.8
$$
it is immediate  to check  that $\Psi$ solves (A.2).

It can  be easily verified that 
the back-rescaled local version near  $a_l, b_l$ matches with the solution
of the outer problem (as $h \to 0$). Thus the issue of the small crosses is settled.

\bigskip

APPENDIX 2. THE VARIATIONAL PROBLEM AND THE FINITE GENUS ANSATZ.

\bigskip

The function $g$ defined by (18) and the conditions stated before 
(18) is crucial for the asymptotic analysis of the
Riemann-Hilbert problem (11). As stated in [KMM] and [KR] the existence of such a function
follows from the existence and regularity of a solution to a variational problem.
In this section we pose the variational problem and we state the results of
[KR] on existence. We also show that a variation of the proofs of [KR]
guarantees the validity of the finite genus ansatz for the barrier data problem.

Let
$ \Bbb H = \{ z: Im z >0 \} $, be the complex upper-half plane  and
 $\bar \Bbb H =  \{ z: Im z \geq 0 \} \cup \{\infty\}$
 be the closure  of $ \Bbb H $. Let
also
$ \Bbb K = \{ z: Im z >0 \} \setminus \{ z: Rez =0, 0< Im z \leq A \}$,
where $A$ is a positive constant.
In the closure of this space, $\bar \Bbb K $, we consider the points
$ix_+$ and $ix_-$, where $0 \leq x < A$ as distinct.
In other words, we cut a slit in the upper half-plane along the
segment $(0, iA)$ and distinguish between the two sides of the slit.
The point infinity belongs to $\bar \Bbb K$, but not $\Bbb K$.
Define
$G(z; \eta)$ to be the Green's function for the upper half-plane
$$
\aligned
G(z; \eta) = log {{ |z-\eta^*| } \over {|z-\eta|}}
\endaligned
$$
and  let $d\mu^0 (\eta)$ be the nonnegative measure $-\rho^0 d\eta$
on the segment $[0,iA]$ oriented from 0 to iA, where
$\rho^0$ is the density of eigenvalues given by  
$$
\aligned
\rho^0 (\lambda) = {{\lambda } \over{\pi   (A^2+\lambda^2)^{1/2} }}.
\endaligned
\tag9a
$$
The star denotes
complex conjugation. Let the "external field" $\phi$ be defined by
$$
\aligned
\phi (z) =
-\int G(z; \eta) d\mu^0(\eta) - Re (i \pi \int^{iA}_z \rho^0 d\eta  +2i  (z x + z^2 t) ),
\endaligned
\tag A.9
$$
where, without loss of generality $x >0$.

Let $\Bbb M$ be  the set of all positive Borel measures on $\bar \Bbb K$,
such  that both the free energy
$$
\aligned
E(\mu) = \int \int G(x,y) d\mu(x) d\mu(y), ~~~\mu \in \Bbb M
\endaligned
$$
and $\int \phi d\mu$ are finite.
Also, let
$$
\aligned
V^{\mu} (z) = \int G(z,x) d\mu(x), ~~~\mu \in \Bbb M.
\endaligned
$$
be the Green's potential of the measure $\mu$.

The weighted energy of the field $\phi$ is
$$
\aligned
E_{\phi} (\mu) =  E(\mu) + 2 \int \phi d\mu, ~~~ \mu \in \Bbb M.
\endaligned
$$

Now, given any curve $F$ in $\bar \Bbb K$, the equilibrium measure
$\lambda^F$ supported in $F$ is defined by
$$
\aligned
E_{\phi} (\lambda^F) = min_{\mu \in  M(F)} E_{\phi} (\mu),
\endaligned
$$
where $M(F)$ is the set of measures in $\Bbb M$ which are supported in $F$,
provided such a measure exists.

The finite gap ansatz is equivalent to the existence of a so-called  S-curve
joining the points $0_+$ and $0_-$ and lying entirely in $\bar \Bbb K $.
By S-curve we mean  an oriented  curve $F$ such that the equilibrium measure
$\lambda^F$ exists, its support consists of a finite
union of analytic arcs
and
at any interior point of $supp\mu$ the so called S-property is satisfied
$$
\aligned
{d \over {d n_+}} (\phi + V^{\lambda^F}) =
{d \over {d n_-}}  (\phi + V^{\lambda^F}).
\endaligned
\tag A.10
$$
In [KR] we show that there is a $C$ such that
$$
\aligned
E_{\phi} (\lambda^C)=
max_{contoursF} E_{\phi} (\lambda^F) = max_{contoursF} min_{\mu \in  M(F)} E_{\phi} (\mu),
\endaligned
\tag A.11 
$$
and that
the existence of an S-curve follows from the existence of a contour $C$ maximizing the
equilibrium measure.

EXISTENCE THEOREM [KR]. For the external field given by (A.9),
there exists a continuum  $F \in \Bbb F$ such that the equilibrium measure
$\lambda^F$ exists and
$$
\aligned
E_{\phi} [F] (= E_{\phi} (\lambda^F)) =
max_{F \in \Bbb F} min_{\mu \in M(F)} E_{\phi} (\mu).
\endaligned
\tag A.12
$$

PROOF: See Theorem 4 in [KR]. Even though the density $\rho$ does not
satisfy all of conditions (1) of Theorem 4, the proof goes through unaltered.

For our particular problem we also have

REGULARITY THEOREM. Under the extra assumption that the continuum $F$
does not touch the spike $\{ z: Rez =0, 0< Im z \leq A \}$ except at 
a finite number of points,
the continuum $F$  is at worst a union of an  S-curve
and a finite union of real intervals.

PROOF: The proof follows as in Theorem 8 of [KR]. 
But there are some changes here.
The density of eigenvalues given by (9a) does not satisfy all
conditions (1) set in [KR]. In particular it is not true that
$Im [\rho^0(z)] >0,~~for~~~z \in (0,iA] \cup \Bbb R^+.$ 
Rather $Im [\rho^0(z)] =0$ on the real line. 
So a small amendment of the regularity proof is needed.

The point of the assumption  that $Im [\rho^0(z)] >0,~~for~~~z \in (0,iA] \cup \Bbb R^+$
is to ensure that the continuum $F$ does not touch the negative real line,
except of course at $0_-$ and possibly infinity. In our case, we can argue as follows.

Because of the
analyticity properties of the field, the  real line can be divided into a finite
number of intervals $J_k$ such that in the interior of each $J_k$ either
$$
\aligned
{{d\phi }\over {dImz}}  >0,
\endaligned
$$
or
$$
\aligned
{{d\phi }\over {dImz}}  \leq 0.
\endaligned
$$
In the first case, one can see that the continuum $F$ cannot touch the real line
except of course at the endpoints of $J_k$.
This is because for any configuration that involves a
continuum  including other points on the real line, we can find a configuration
with no other points on the real line, by pushing
measures up away from the real axis, which  has
greater  (unweighted $and$ weighted) energy. 
It is crucial here that if $u \in \Bbb R$ then
$G(u,v)=0$, while if both $u, v $ are off the real line
$G(u,v) > 0.$  

In the second case, we get a finite union of real intervals.

So $F$ consists of a finite union of arcs: some of them are real intervals
(at worst) and the rest do not touch the  real line except at
their endpoints.

To pursue the proof of regularity
one neeeds the following identity.

THEOREM [KR]. Let $F$ be the maximizing continuum of
and $\lambda^F$ be the  equilibrium measure.
Let $\mu$ be the extension of $\lambda^F$ to the
lower complex plane via $\mu(z^*) = -\mu(z)$.
Then
$$
\aligned
Re (\int {{d\mu(u)} \over {u-z}} + V'(z) )^2=
Re ( V'(z))^2 - 2 Re \int {{V'(z)-V'(u)} \over {z-u}} d\mu(u)  \\
+ Re [ {1 \over z^2}  \int 2 (u+z) V'(u) ~ d\mu(u)  ] .
\endaligned
\tag A.13 
$$
Here $V$ is the complexification of the real field $\phi$.

PROOF: By taking variations with respect to the equilibrium measure;
see Theorem 5 of [KR]. Of course, variations cannot be taken near
the real intervals, but since the field $\phi =0$ and the Green's function is
also zero, these real intervals  are  not  
part of the support of the equilibrium measure.

From (A.13) it is easy to see that the support of the equilibrium measure
of the maximizing continuum is characterized by
$$
\aligned
Re \int^z (R_{\mu})^{1/2} dz =0,
\endaligned
\tag A.14
$$
where
$$
\aligned
R_{\mu}(z)=
( V'(z))^2 - 2  \int_{supp\mu} {{V'(z)-V'(u)} \over {z-u}} d\mu(u)  \\
+ {1 \over z^2} (\int_{supp\mu} 2 (u+z) V'(u) ~ d\mu(u) ) .
\endaligned
$$
Since $ R_{\mu}(z)$
is a  function analytic in $\Bbb K$,
the locus defined by (A.14) is a union of arcs with endpoints at zeros of
$R_{\mu}$. But it is readily seen  that $R_{\mu}$ has finitely many zeros.

(A.10) also follows easily from (A.13); see [KR].
Alternatively, see Chapter 8 of [KMM].

\bigskip

If $F$ touches the spike $[0,iA]$ at more than a finite number of points,
regularity cannot be proved as above because variations
cannot be taken. In [KR] we have included a rough idea
on how to extend the above proof. Here is
a more detailed argument forthcoming (see also [K]).

One wishes to somehow allow the contour $F$ go through the spike $[0,iA]$. One
problem arising is that (the complexification of)
the external field is not analytic across the segment $[-iA, iA]$.
What is true, however, is that $V$ is analytic in a Riemann surface consisting
of infinitely
many sheets, cut along the line segment
$[-iA, iA]$. So, the appropriate underlying space for the
(doubled up) variational problem should now be a
non-compact Riemann surface, say $\Bbb L$. Now,
compactness is the crucial element in the proof
of a maximizing continuum. But we can
indeed compactify the
Riemann surface $\Bbb L$ by mapping it to a subset of the complex plane
and compactifying the complex plane.
The other problem, of course, is whether the amended variational problem
(with the modified
field defined on the Riemann surface
and with the possibility of $F$ not enclosing all
the original eigenvalues) is still appropriate
for the semiclassical  NLS.
The argument  goes roughly as follows:

(i) Proof of the existence of an S-curve $F$ in $\Bbb L$
along the lines of [KR].

(ii) Deformation of the original discrete Riemann-Hilbert problem to the set
$\hat F$ consisting of the  projection of  $F$
to the complex plane.
At first sight, it is clear that  $\hat F$ may not encircle the spike $[0,iA]$.
It is however possible  to append S-loops (not necessarily with respect to the same
branch of the external field)
and end up with a sum of S-loops, such that the amended $\hat F$  $does$
encircle the spike $[0,iA]$. To see this,  suppose
there is an open interval, say $(i\alpha, i\alpha_1)$,
which lies in the exterior of $\hat F$, while
$i\alpha, i\alpha_1 \in \hat F$. Let us assume, that $\hat F$ crosses
$[0,iA]$ along bands at $i\alpha, i\alpha_1$ (if not the situation is
similar and simpler); call these bands $S, S_1$.
Let $\beta^-, \beta^+$ be points (considered in $\Bbb C$)
lying on $S$ to the left and right
of $i\alpha$ respectively, and at a small distance
from $i\alpha$. Similarly,
let $\beta_1^-, \beta_1^+$ be points lying on $S_{1}$
to the left and right
of $i\alpha_{1}$ respectively, and at a small distance
from $i\alpha_{1}$.
We will show that there exists a "gap" region including the preimages of
$\beta^-, \beta_{1}^-$ lying in the $N$th sheet for $-N$ large enough,
and similarly there exists a "gap" region including the preimages of
$\beta^+, \beta_{1}^+$ lying in the $M$th sheet for $M$ large enough,
both being  regions for which the gap inequalities hold a priori,
irrespectively of the actual S-curve,
depending only on the external field!

Indeed, note  that the
quantity  $Re (\tilde \phi^{\sigma} (z) )$ (which defines the variational
inequalities) is a priori bounded above  by
$-\phi(z)$. For this, see (8.8) in Chapter 8 of [KMM]; there is actually a sign error:
the right formula is
$$
\aligned
Re (\tilde \phi^{\sigma} (z) )= -\phi(z)
+ \int G(z,\eta) \rho^{\sigma} (\eta)d\eta.
\endaligned
$$
Next note that the difference of the values of the
function $Re (\tilde \phi^{\sigma} (z) )$ in consecutive sheets is
$\delta Re (\tilde \phi^{\sigma}) = \pm 2 \pi Rez$, and hence the difference of
the values at points on consecutive sheets whose image under the
projection to the complex plane is $i \eta + \epsilon$, where $\eta $ is real
and $\epsilon $ is a small (negative or positive) real,
is $\delta (Re \tilde \phi^{\sigma} )= \pm 2 \pi \epsilon$.
This means that on the left (respectively right)
side of the imaginary semiaxis, the inequality $Re ( \tilde \phi^{\sigma} (z) )<0$
will be eventually (depending
on the sheet) be valid at any given small distance to it.

Applying the theory of [KR] we join the preimages of $\beta^-$
and  $\beta_{1}^-$ (under the projection pf $\Bbb L $ to
$\Bbb C$) lying in the $N$th sheet and the preimages of $\beta^+$
and  $\beta_{1}^+$ lying in the $M$the sheet.
Finally we connect the  preimage of $\beta^+$
lying in the $N$th sheet to the preimage of $\beta^-$
lying in the $M$th sheet,  and so on. 
We thus end up with an S-loop whose projection is
covering the "lacuna" $(i\alpha, i\alpha_{1})$.

The original discrete Riemann-Hilbert problem  can be  trivially deformed to a
discrete Riemann-Hilbert on the resulting  (projection of the)
union of S-loops. All this is possible even in the
case where $\hat F$ self-intersects.

(iv) Deform the discrete Riemann-Hilbert problem to the continuous one with the right
band/gap structure
(on $\hat F$; according to the projection of the  equilibrium measure on $F$), which is
then explicitly solvable via theta functions.
Both the discrete-to-continuous approximation and the opening of the lenses needed
for this deformation are justified  as in [KMM] (see also the article [LM] 
for the delicate study of the Riemann-Hilbert problem near the points where
$\hat F$ crosses the spike).
The g-function is defined by the same Thouless-type formula with respect to the
equilibrium measure (cf. section 2(iii)).
It satisfies the same conditions as in [KMM]
(measure reality and variational inequality) on bands
(where the branch of the field turns out to be irrelevant) and
on gaps (where the inequalities  are
satisfied according to the branch of the external field).

We arrive at the following conclusion.

FINITE GAP ANSATZ. The finite genus ansatz is valid generically in $x,t$  for the steplike data
given by (2).

\bigskip

BIBLIOGRAPHY
\bigskip

[BK] J.C.Bronski,  J.N.Kutz,
Numerical Simulation of the Semi-Classical Limit of the Focusing Nonlinear 
Schr\"odinger Equation, Physics Letters A, v.254,  no.6,  1999,  pp.325-336.

[C] H.Ceniceros, A Semi-Implicit Moving Mesh Method for the Focusing Nonlinear 
Schr\"odinger Equation,  Commun. Pure Appl. Anal. v.1, no. 1, 2002, pp.1--18.

[CK] A.Cohen, T.Kappeler, Solutions to the Cubic Schr\"odinger Equation
by the Inverse Scattering Method, SIAM Journal of  Mathematical Analysis,
v.23, n.4, 1992, pp.900-922.

[CT]  H.Ceniceros,  F.R.Tian,  A
Numerical Study of the Semi-Classical Limit of the Focusing Nonlinear 
Schr\"odinger Equation, Phys. Lett. A, v.306, no.1, 2002, pp.25--34. 

[DVZ1] P.Deift, S.Venakides, X.Zhou,
The Collisionless Shock Region for the Long-Time
Behavior of Solutions of the KdV equation, Communications in Pure and Applied
Mathematics, v.47, 1994, pp.199-206.

[DVZ2] P.Deift, S.Venakides, X.Zhou,
New Results in Small Dispersion  KdV  by an Extension of the Steepest
Descent Method for Riemann-Hilbert Problems, IMRN 1997, pp.286-299.

[DZ1] P.Deift, X.Zhou, A Steepest Descent Method for Oscillatory Riemann-Hilbert
Problems, Annals of Mathematics, 2nd Series, Vol. 137, 
No. 2., 1993, pp. 295-368.

[DZ2]  P.Deift, X.Zhou, Long Time asymptotics for Solutions of the NLS
Equation with Initial Data in a Weighted Sobolev Sppace,
Communications in Pure and Applied
Mathematics, v.56, 2003, pp.1029-1077.
 
[K] S.Kamvissis, From Stationary Phase to Steepest Descent, invited   
contribution to a volume honoring P.Deift, Contemporary Mathematics, v.458,
AMS 2008.

[KMM] S.Kamvissis, K.McLaughlin,  P.Miller, Semiclassical Soliton Ensembles 
for the Focusing Nonlinear Schr\"odinger Equation, 
Annals of Mathematics Studies, v.154, Princeton University Press, 2003;
arXiv:nlin/0012034.

[KR] S.Kamvissis, E.Rakhmanov, Existence and Regularity for an
Energy Maximization Problem in Two Dimensions, 
Journal of Mathematical Physics, v.46, n.8, 2005.

[LM] G.Lyng, P.Miller, The N-soliton of the focusing nonlinear Schr\"odinger
equation for N large, Comm. Pure Appl. Math., v.60, 2007, pp. 951-1026.

[MK]  P.Miller, S.Kamvissis, On the Semiclassical Limit of the Focusing 
Nonlinear Schr\"odinger 
Equation, Phys. Lett. A, v.247, no. 1-2, 1998, pp.75--86. 

[T] Fei-Ran Tian, private communication.

[TVZ] A.Tovbis, S.Venakides, X.Zhou, Communications in Pure and Applied
Mathematics, v.57, 2004.

\enddocument